\documentclass[12pt]{article}
\newlength{\abstractwidth}
\input epsf
\usepackage{epsfig}

\makeatletter
\newcounter{fig}
\renewcommand\thefig{\arabic{fig}}

\def\fps@fig{tbp}
\def\ftype@fig{1}
\def\ext@fig{lof}
\def\fnum@fig{\figurename~\thefig}

\newenvironment{fig*}
               {\@dblfloat{fig}}
               {\end@dblfloat}

\renewcommand{\thefootnote}{\fnsymbol{footnote}}
\renewcommand{\thanks}[1]{\footnote{#1}}
\newcommand{\starttext}{
\setcounter{footnote}{0}
\renewcommand{\thefootnote}{\arabic{footnote}}}

\newcommand{\bea}{\begin{eqnarray}}
\newcommand{\eea}{\end{eqnarray}}
\newcommand{\ee}{\end{equation}}
\newcommand{\be}{\begin{equation}}
\newcommand{\<}{\langle}
\renewcommand{\>}{\rangle}

\def\A{{\cal A}}
\def\B{{\cal B}}

\def\D{{\cal D}}

\def\G{{\cal G}}

\def\L{{\cal L}}
\def\M{{\cal M}}

\def\O{{\cal O}}

\def\R{{\cal R}}

\def\W{{\cal W}}

\def\Y{{\cal Y}}

\def\bC{{\bf C}}
\def\bG{{\bf G}}
\def\bZ{{\bf Z}}

\def\Re{{\rm Re}}
\def\Im{{\rm Im}}
\def\tr{{\rm tr}}
\def\det{{\rm det}}
\def\Det{{\rm det}}

\def\half{ {1\over 2}}
\def\z{{\bf z}}

\def\p{\partial}

\def\tet{\vartheta}
\def\ep{\varepsilon}
\def\e{\epsilon}
\def\o{\omega}
\def\v{\varpi}
\def\f{\varphi}

\def\no{\nonumber}

\begin{document}
\starttext
\baselineskip=16pt
\setcounter{footnote}{0}

\begin{flushright}
UCLA/05/TEP/08 \\
Columbia/Math/05 \\
2005 March 22\\
\end{flushright}

\bigskip

\begin{center}
{\Large\bf TWO-LOOP SUPERSTRINGS  AND S-DUALITY} 
\footnote{Research supported in part by National Science
Foundation grants PHY-01-40151, PHY-02-45096, and DMS-02-45371.}

\bigskip\bigskip
{\large Eric D'Hoker$^a$, Michael Gutperle,$^a$ and D.H. Phong$^b$}

\bigskip
$^a$ \sl Department of Physics and Astronomy \\
\sl University of California, Los Angeles, CA 90095, USA\\
$^b$ \sl Department of Mathematics\\
\sl Columbia University, New York, NY 10027, USA

\end{center}

\bigskip\bigskip

\begin{abstract}

\end{abstract}

The two-loop contribution to the Type IIB low energy effective action term 
$D^4 R^4$, predicted by SL(2,Z) duality, is compared with that of
the two-loop 4-point function derived recently in superstring perturbation
theory through the method of projection onto super period matrices. 
For this, the precise overall normalization of the 4-point function is
determined through factorization. The resulting contributions to $D^4
R^4$ match exactly,  
thus providing an indirect check of SL(2,Z) duality. The two-loop Heterotic 
low energy term $D^2F^4$ is  evaluated in string perturbation theory;
its form is closely related to  the $D^4 R^4$ term in Type II,
although its significance to  duality is  an open issue.

\vfill\eject

\baselineskip=15pt
\setcounter{equation}{0}
\setcounter{footnote}{0}

\newpage

\section{Introduction}
\setcounter{equation}{0}

The dualities \cite{Hull:1994ys,Witten:1995zh} between various superstring 
theories and M theory provide strong constraints on their effective actions.  
In \cite{Green:1997tv} it was conjectured that the $SL(2,{\bf Z})$ duality 
(or S-duality) of Type IIB string theory completely determines the $R^4$ 
term in the  superstring effective action. 
The dependence of the $R^4$ term on the
axion/dilaton scalar field $\tau=\chi+ie^{-\phi}$ is given by a
non-holomorphic Eisenstein series  (or Maass waveform \cite{Terras}) 
$E_{3/2}(\tau,\bar \tau)$. A remarkable consequence of this
conjecture is that the $R^4$ term receives contributions only 
from tree-level and one-loop orders in string perturbation theory
(in addition there are non-perturbative D-instanton contributions).

\medskip

Further evidence for this conjecture was provided in 
\cite{Green:1997di,Green:1997as}, where the $R^4$ term was derived
from a one-loop amplitude of eleven dimensional supergravity. In
\cite{Green:1998by} it was shown that the exact form of the $R^4$ and
related terms \cite{Green:1999by,deHaro:2002vk} is completely determined by
supersymmetry and S-duality. Utilizing U-duality symmetries, these 
results for the $R^4$ term were extended to toroidally compactified 
Type II string theory  in
\cite{Pioline:1998mn,Kiritsis:1997em,Obers:1999es,Obers:1999um,Berkovits:1997pj}. Related conjectures for $R^4 H^{4g-4}$ terms were discussed in
\cite{Berkovits:1998ex}.
 
 \medskip

The $R^4$ term is expected to receive no renormalizations beyond one-loop, 
since it can be expressed as an integral over half the (linearized and on-shell) 
Type IIB superspace,  and thus is protected by supersymmetry.  
Dimensional arguments suggest that terms with additional  derivatives should 
enjoy a similar protection.  In \cite{Green:1999pu,Green:1999pv},
a two-loop calculation in eleven dimensional supergravity was used to 
show that  the axion/dilaton  dependence of $D^4 R^4$ term in the superstring effective action is governed by a non-holomorphic
Eisenstein series $E_{5/2}(\tau,\bar \tau)$. The functional form of the
Eisenstein series is such that it generates only tree-level and two-loop 
contributions in string perturbation theory to this term. 

\medskip

Recently, a systematic method for constructing the superstring $N$-point
function to two-loops from first principles has been developed \cite{I,IV,V,VI} 
(see also \cite{dp02} for a review). With this method, we have a 
completely explicit expression for the superstring 4-point function
of massless NS bosons \cite{VI,zwzIII}. In particular, this construction 
provides a first principles proof that the two-loop amplitude does not 
contribute to the $R^4$ term in the effective action  \cite{VI}. 
This result may be interpreted as an indirect check of the 
conjectured $SL(2,{\bf Z})$ duality structure of the $ R^4$ terms.
The vanishing of the two-loop contribution to the $ R^4$ terms had also
been argued earlier by many authors \cite{bis3, lech, iz}. However, the
calculations they had to rely  on were gauge-slice dependent and did not 
provide  reliable checks.

\medskip

The purpose of the present paper is to compare the contribution to the 
$D^4 R^4$ term deduced from the two-loop superstring amplitude \cite{VI, zwzIII} 
to the one predicted by \cite{Green:1999pu} on the basis of $SL(2,{\bf Z})$ 
duality. Since we shall be comparing two non-vanishing contributions, their overall normalizations will  have to  be determined with particular care. Actually, 
the two-loop 4-point function was determined in \cite{VI, zwzIII} only up to an
overall constant,  left undetermined due to the occurrence of an unknown 
overall normalization constant in the  chiral bosonization formulas of \cite{vv}. 
Several attempts have been made at calculating these constants, 
but even just for the $bc$ ghost spin 1 bosonization formula,
no generally agreed upon values seems to have been attained as yet \cite{jorg}.

\medskip

Precise overall normalization constants are of critical importance for 
the detailed comparison between results from duality and superstring
perturbation theory that we shall carry out in this paper. 
One of our main tasks here is to  determine precisely the overall 
normalization of the two-loop 4-point function. We accomplish this through 
factorization  constraints to tree-level and one-loop. Many different
conventions used in the literature, as well as the occurrence of an 
occasional typo, have forced us to carefully rederive also the basic tree-level 
and one-loop amplitudes. A systematic account is provided in Appendices B and C.

\medskip

With these precise overall normalizations,
we find  complete agreement between the two-loop perturbative value for the 
$D^4 R^4$ term and the value predicted by $SL(2,{\bf Z})$ duality, thus 
providing another indirect check for S-duality.

\medskip

Finally, the Heterotic strings \cite{Gross:1985fr} inherit  half of the 
supersymmetry  and some of the non-renormalization  effects of 
Type II theories in perturbation theory, even though their non-perturbative 
structure is very different from that of Type II strings. 
In particular, the $Spin(32)/\bZ_2$ Heterotic string is expected to be 
dual to the  Type I string \cite{Witten:1995ex, Tseytlin:1995fy}, 
(for a review, see \cite{Bachas:1998rg}).
In \cite{VI}, it was shown that in the low  energy effective action for the 
Heterotic  strings, terms in $F^4$, $F^2F^2$,  $R^2F^2$ and $R^4$ 
receive no two-loop contributions. While the non-renormalization  of the 
$F^4$ and $F^2F^2$  terms had been argued earlier in \cite{STI},  that
of the $R^4$ terms appears to be at odds with an  earlier duality 
analysis \cite{Tseytlin:1995fy}.  To clarify these issues, it may be helpful to 
better understand the renormalization properties of terms that involve also additional derivatives. In this paper, we calculate the two-loop contribution
to the  $D^2 F^2 F^2$ term, and find that its magnitude involves a remarkable 
modular form. The interplay between this result and S-duality is an issue 
left for later study.

\medskip

The remainder of the paper is organized as follows. In section 2, the 
predictions for the Type IIB low energy effective action terms $R^4$
and $D^4 R^4$, both from S-duality and from perturbation theory
are summarized, compared and found to be in perfect agreement.
The precise normalization of the two-loop massless NS-NS 4-point 
function in Type II  is obtained by factorization onto one-loop
components in sections 3. From this 4-point function, the perturbative
contribution to the $D^4 R^4$ term is derived in section 4. The 
analogous calculation of the two-loop corrections to $D^2F^2F^2$ terms 
in the Heterotic string is provided in section 5. In Appendix A, 
basics of genus 1 and 2 moduli spaces are summarized.
Careful calculations of the normalization of the 4-point functions
are given to  tree-level in Appendix B and to one-loop
in Appendix C. Regulator dependence is discussed in Appendix D.

\newpage

\section{Summary of S-Duality and Perturbative Results}
\setcounter{equation}{0}

In this section, we collect the predictions of S-duality for the 
$R^4$ and $D^4 R^4$ terms in the low energy effective action of Type IIB, 
and express them in a notation that will facilitate their comparison with the 
results from perturbation theory. The latter are summarized in this section, 
but their derivation is postponed until sections 3 and~4. A detailed comparison  
between both sets of predictions reveals perfect agreement.

\subsection{S-duality prediction for $R^4$ and $D^4 R^4$ terms }

The massless spectrum of Type IIB string theory contains two scalar
fields, the NS-NS dilaton $\phi$ and the R-R axion $\chi$. The
theory has a remarkable non-perturbative $SL(2,\bZ )$ S-duality symmetry
under which a complex combination $\tau= \chi+ie^{-\phi}$ transforms as
\be
\tau\to {a\tau+b\over c\tau+d}, \hskip 1in a,b,c,d\in \bZ, \quad ad-bc=1
\label{modtra}
\ee
In the Einstein frame, the action for the graviton is given by 
\be
S= {1\over 2 \kappa_{10}^2} \int d^{10}x \sqrt{-G_E} \, R _E
\ee
where the superscript $E $ indicated that the metric $G_E$ and the 
Ricci scalar $R_E$ are expressed in the Einstein frame.   S-duality  acts 
naturally in the Einstein frame where the metric is invariant  under the
transformation (\ref{modtra}).  
Therefore, it is expected that all terms in the effective action which involve
only the metric and the dilaton/axion fields should have a dependence on 
$\tau$ and $\bar \tau$ which is invariant under (\ref{modtra}).

\medskip

The Einstein frame is not the natural frame for expressing the coupling
of the  string sigma model to the space time metric. Instead, one uses
the string frame metric $G$, which is related to the Einstein frame metric by
\be
G _{\mu \nu} =e^{\phi/2} G_{E \mu \nu}
\ee 
The gravitational part of the action  in the string frame becomes
\bea
S = { 1 \over 2 \kappa _{10}^2} \int d^{10}x ~ \sqrt{-G} \, e^{-2 \phi} \, R
\eea
where $R$ is now the Ricci scalar for the metric $G$.
Henceforth, we shall use the string frame metric throughout.
Notice that in the Einstein-Hilbert action, only the combination $\kappa _{10}^2 e^{2 \phi}$ enters, a combination that is unaffected by a shift in $\phi$
compensated by a multiplicative factor in $\kappa _{10}^2$. The normalization
chosen here for $\phi$ is such that the S-duality transformation
law for $\tau = \chi + i e^{-\phi}$ is canonical, as in (\ref{modtra}).

\medskip

It was conjectured in \cite{Green:1997tv} that the $R^4$ terms in the  
Type IIB effective action take the following form, (in the string frame),
\bea
\label{SR4}
S_{R^4}=C_{R^4}\int d^{10}x\,\sqrt{-G} \, \R^4\,e^{-{1\over 2}\phi}\, 
2\zeta(3)E_{3/2}(\tau,\bar\tau)
\label{rfourterm}
\eea
The expression $\R^4$ stands for $\R^4=t_8t_8R^4$, where $t_8$ is the 
well-known kinematic tensor which enters both the  tree-level and one-loop 
superstring 4-point functions and is defined here as it was in \cite{gs,GSW}.
The overall constant $C_{R^4}$ will be discussed later.
Furthermore, $\zeta (s)$ is the Riemann zeta function, and 
$E_{3/2}(\tau,\bar\tau)$ is the non-holomorphic Eisenstein
series of weight $s=3/2$. For general $s$, the non-holomorphic
Eisenstein series  $E_{s}$ is defined  by 
\bea
 2\zeta(2s) E_s (\tau, \bar \tau) = \sum_{(m,n)\neq (0,0)}
 {\tau_2^s\over |m+n \tau|^{2s}} 
\label{einsser}
\eea
and satisfies  the following  differential equation
\bea
4\tau_2^2 \partial_\tau \partial_{\bar\tau} E_{s}(\tau,\bar\tau)=
s(s-1) E_{s}(\tau,\bar\tau)
\label{lapeq} 
\eea
In \cite{Green:1998by} it was shown that (\ref{lapeq}) is  a
consequence of supersymmetry. It is easy to see that (\ref{lapeq})
has two  solutions of the form $c_{-s}\tau_2^{s}$ and
$c_{s-1}\tau_2^{1-s}$ corresponding to two particular orders of perturbation
theory. The series expression (\ref{einsser}) is the unique S-duality
invariant solution with these perturbative terms.
In \cite{Green:1997as} the form of the $R^4$ term (\ref{SR4}) was derived 
from a one-loop calculation in eleven dimensional supergravity.
The expansion of the non-holomorphic Eisenstein series
$2\zeta(3)E_{3/2}(\tau,\bar\tau)$ is given in \cite{Terras}
\bea
\label{E32}
2\zeta(3)E_{3/2}(\tau,\bar\tau)
=2\zeta(3)e^{-{3\over 2}\phi}+{2\pi^2\over 3}e^{{1\over 2}\phi}
+
{\rm non-perturbative}
\eea
It follows that the $R^4$ term in the effective action  (\ref{rfourterm}) gets
perturbative contributions only from tree-level and one-loop. 
The vanishing of two-loop contributions was proven in \cite{VI}.

\medskip

The $R^4$ term may be expressed as an integral over half the
(linearized on-shell) IIB superspace and should therefore be protected by
supersymmetry. Dimensional analysis suggests that terms with up to six
derivatives acting on $R^4$ should still be protected by supersymmetry.
In \cite{Green:1999pu} a two-loop calculation in eleven dimensional 
supergravity was used to calculate the $D^4 R^4$ terms explicitly  in
the Type IIB effective action. It was found that  it has a S-duality invariant 
form (still expressed in the string frame), 
\bea
S_{D^4 R^4} = C_{D^4R^4}\,\int d^{10} x \sqrt{ -G} \, D^4 \R^4 \, e^{{1\over 2} \phi}
\, \zeta(5)\,E_{5/2}(\tau,\bar \tau) 
\label{greencon}
\eea
For $E_{5/2}$ the expansion in large negative $\phi$ results in two perturbative terms,  given by
\bea
\label{E52}
2\zeta(5) E_{5/2} = 2 \zeta(5)e^{-{5\over 2} \phi} + {\pi^4 \over 90}\times
{8\over 3} \; e^{{3\over 2}\phi}+ {\rm non-perturbative}
\eea
From (\ref{greencon}) it is clear that the two terms come from a tree-level 
and a two-loop contribution, but that the one-loop  contribution is absent. 
This is in accord with \cite{Green:1999pv}, where the 
one-loop contribution to this term in the action was shown to be zero.

\medskip

The  normalization constants $C_{R^4}$ in (\ref{SR4}) and $C_{D^4R^4}$
in (\ref{greencon}) can be determined from the tree-level 4-point function\footnote{Throughout, we shall omit the overall momentum 
conservation factor $(2\pi)^{10} \delta ^{(10)} (k_1+k_2+k_3+k_4)$ 
when expressing the scattering amplitudes, and use the Mandelstam variables, $s=- (k_1+k_2)^2$, $t =- (k_2+k_3)^2$, and $u = - (k_1+k_3)^2$. }
\bea
\label{tree0}
A^{(4)} _0
= \R^4 \kappa_{10}^2 e^{-2\phi} 
{\Gamma(-s \alpha' /4 ) \Gamma(- t \alpha' / 4) \Gamma(- u \alpha' / 4 )
\over 
\Gamma(1+ s \alpha' /4 ) \Gamma(1+ t \alpha' / 4)
\Gamma(1+ u\alpha' / 4)}
\eea
as follows. We use the formula
\bea
{\Gamma(1-z)\over \Gamma(1+z)} = e^{2\gamma z} \exp \left (
\sum_{k=1}^\infty  {2\zeta(2k+1) z^{2k+1} \over 2k+1} \right )
\eea
to derive the expansion of the 4-point function in powers of $s$, $t$ and $u$,
\bea
A^{(4)} _0
= \R^4 \kappa_{10}^2 e^{-2\phi} {2^6\over (\alpha')^3
  stu}\exp \left ( \sum_{k=1}^\infty {2\zeta(2k+1)\over
  2k+1} ( \alpha' /4 )^{2k+1}( s^{2k+1}+t^{2k+1}+u^{2k+1}) \right )
\eea
Expanding to the order needed for the study of $R^4$ and $D^4 R^4$ terms
yields,
\bea
A^{(4)} _0
= \R^4 \kappa_{10}^2 e^{-2\phi} \left ( {2^6 \over (\alpha')^3
  stu}  + 2 \zeta(3)+{\zeta(5) \over 16}(\alpha')^2 ( s^2+t^2+u^2) \right )
\label{treeex}
\eea
The first term in (\ref{treeex}) arises through 1-particle reducible Feynman 
diagrams from the Einstein-Hilbert action.
The second term in (\ref{treeex}) gives the following tree-level contribution 
to the $R^4$ terms in the effective action \cite{Gross:1986iv}
\bea
A_0^{(R^4)}=\R^4  \kappa_{10}^2 2 \zeta(3)e^{-2\phi}
\eea
Now the $R^4$ terms in the effective action are given
by the formula (\ref{SR4}) in terms of $2\zeta(3)E_{3/2}(\tau,\bar\tau)$. 
The expansion (\ref{E32}) for $2\zeta(3)E_{3/2}(\tau,\bar\tau)$ gives 
then the relative coefficients between the tree-level and the one-loop
contributions to the $R^4$ terms in the effective action.
It follows that the normalization of the one-loop contributions 
to the $R^4$ terms corresponding to the normalization
(\ref{treeex}) for the 4-point function is given by
\bea
\label{A1R4}
A_1^{(R^4)}=\R^4\kappa_{10}^2{2\pi^2\over 3}
\eea
The third term in (\ref{treeex}) gives the following tree-level contribution
to the $D^4 R^4$ terms,
\bea
A_0^{(D^4R^4)}=\R^4 (s^2+t^2+u^2) \kappa_{10}^2 {(\alpha')^2 \over 2^4}
\zeta(5)e^{-2\phi}
\eea
The relative coefficients of the tree-level and the two-loop
contribution to the $D^4R^4$ term can be read off from the
expansion (\ref{E52}) of $2\zeta(5)E_{5/2}(\tau,\bar\tau)$.
It follows that the two-loop contribution must be given by
\bea
\label{D4R41}
A_2^{(D^4R^4)}
=
\R^4 \kappa_{10}^2{(\alpha')^2\over 2^4}{4\pi^4\over 270}e^{2\phi}(s^2+t^2+u^2)
\eea
Note that the one-loop contribution to $D^4R^4$ indeed 
vanishes in view of \cite{Green:1999pv}.

\subsection{Normalizations of Superstring Perturbation Theory}

The starting point for superstring perturbation theory is the 
action $I_m$ and vertex operators for massless NS-NS
states $V$ on a worlsheet $\Sigma$, normalized as follows,
\bea
\label{basics}
I_m 
& = & \lambda \chi (\Sigma) ~ + ~
{1 \over 2 \pi \alpha '} \int _\Sigma d^{2|2} \z E \D_- X^\mu \D_+ X^\mu 
\no \\
V (\epsilon, \bar \epsilon, k) 
& = & 
\kappa \, \ep ^\mu \bar \ep ^\nu \int _\Sigma d^{2|2} \z E  
\D_- X^\mu \D_+ X^\nu  e^{i k \cdot X}
\eea
Here, $\chi (\Sigma) = 2-2h$ is the Euler number for a closed surface
$\Sigma$ with $h$ handles, and $\lambda$ is the corresponding coupling
constant\footnote{While $\lambda$ and the vacuum expectation 
value of the dilaton $\phi$ are clearly related objects, their 
precise normalizations may differ by an additive constant (see Appendix D), 
and the form of S-duality transformations, expressed in terms of $\lambda$, 
may not be  canonical as in (\ref{modtra}).}
governing the perturbative loop expansion in powers 
of the string coupling $g_s \sim e^{-2 \lambda}$.
Furthermore,  $\z = (z,\theta)$ is the super-coordinate on the worldsheet $\Sigma$, and the superfield $X^\mu$ is given in terms of 
component fields by 
$X^\mu = x^\mu + \theta \psi _+ ^\mu + \bar \theta \psi ^\mu _- 
+ i \theta \bar \theta F^\mu$, where $F$ is an auxiliary field.
Finally, $\kappa$ is the normalization constant of the massless vertex
operators, which is fixed in terms of the other parameters using 
unitarity. Its precise value will not be needed here.

\medskip

The general supergeometry expressions for the various ingredients 
in (\ref{basics}) are given in \cite{dp88,I}.
On a superconformally flat worldsheet, we have $E=1$, and
$\D_+ = \p_\theta  + \theta \p_z$,
$\D_- = \p_ {\bar\theta} + \bar \theta \p_{\bar z}$.
Using the convention $\int d^2 \! \theta \, (\theta \bar \theta) =1$, the above
definitions give us back the standard component expressions for the action
and the vertex operators, 
\bea
\label{basics2}
I_m 
& = & \lambda \chi (\Sigma) ~ + ~
{1 \over 2 \pi \alpha '} \int _\Sigma d^2z
\left ( \p x^\mu \bar \p x^\mu - \psi _+ ^\mu \bar \p \psi _+ ^\mu 
- \psi _- ^\mu \p \psi _- ^\mu \right )
\no \\
V (\epsilon, \bar \epsilon, k) 
& = & 
\kappa \, \ep ^\mu \bar \ep ^\nu
\int _\Sigma d^2 z (\p x^\mu + i k^\rho \psi _+ ^\rho \psi _+ ^\mu)
(\bar \p x^\nu + i k^\sigma  \psi _- ^\sigma \psi _- ^\nu) e^{i k \cdot x}.
\eea
For tree-level and one-loop orders, these expressions are sufficient,
but at higher genus, the supermoduli must be carefully taken into
account as well, and (\ref{basics2}) has additional dependence on
Beltrami differentials and worldsheet gravitini \cite{dp88,V}.

\subsection{Perturbative Predictions}

With the above normalizations and conventions, the Type II superstring 4-point
functions for massless NS-NS bosons  to tree-level, one-loop and two-loop
orders are given as follows. To tree-level, we have\footnote{The amplitudes
predicted from S-duality were denoted by Latin capitals $A$, while the 
amplitudes resulting from superstring perturbation theory will be
denoted by calligraphic capitals $\A$.} 
\bea
\label{tree4}
\A_0 ^{(4)} (\e_i,k_i)
= C_0 \, Q_0 \,    e^{-2\lambda} \, \kappa ^4\, K \bar K \,
{2\pi \Gamma (-\alpha ' s/4 )\Gamma (- \alpha ' t/4)\Gamma (- \alpha ' u/4)
\over
\Gamma (1+\alpha ' s/4)\Gamma (1+\alpha ' t/4)\Gamma (1+\alpha ' u/4)},
\eea
To one-loop, we have\footnote{One-loop moduli are customarily denoted 
by $\tau$ and $\bar \tau$; clearly they are not to be confused with the 
complex dilaton/axion field $\tau= \chi + i e^{-\phi}$.} 
\bea
\label{one4}
\A_1 ^{(4)} (\e_i, k_i)
=
C_1 \,  \kappa ^4 \, K \bar K \, 
\int _{\M_1} {|d\tau|^2 \over ( \Im \, \tau)^2}
\prod _{i=1}^4 \int _\Sigma {d^2z_i \over \Im \, \tau}
\exp \left \{ - {\alpha ' \over 2} \sum _{i<j} k_i \cdot k_j   G(z_i,z_j) \right \}
\eea
and to two-loop order, we have 
\bea
\label{TypeII}
\A_2 ^{(4)} (\epsilon_i, k_i)
=
C_2 \, e^{2 \lambda} \, \kappa ^4 \, K \bar K  \, 
\int_{\M_2} { |d^3 \Omega|^2 \over (\det \Im \Omega)^5}
\int_{\Sigma^4}
|{\cal Y}_S|^2
{\rm exp}\bigg(-{\alpha'\over 2}\sum_{i<j}k_i\cdot k_j\,G(z_i,z_j)\bigg)
\eea
Here $G(z,w)$ are the conformally invariant Green's functions for genus 1 and genus 2,
\bea
\label{scalarprop}
G(z,w)=-\ln |E(z,w)|^2+2\pi ({\rm Im}\,\Omega)_{IJ}^{-1} \,
\biggl ( {\rm Im}\int_z^w\omega_I \biggr )
\biggl ( {\rm Im}\int_z^w\omega_J \biggr )
\eea
The quadri-holomorphic 1-form $\Y_S$ is given by
\bea
\label{YS}
3 \Y_S =
(t-u) \Delta (1,2) \Delta (3,4) +
(s-t) \Delta (1,3) \Delta (4,2) +
(u-s) \Delta (1,4) \Delta (2,3)
\eea
where the basic bi-holomorphic antisymmetric 1-form $\Delta$ is defined by
\bea
\label{Delta}
\Delta (z,w) = 
\omega _1(z) \omega _2 (w) - \omega _1(w) \omega _2 (z).
\eea
The  kinematic factor $K$ is normalized as follows,
\bea
\label{kin4}
K 
& \equiv &
(f_1f_2) (f_3 f_4) + (f_1f_3) (f_2 f_4) + (f_1f_4) (f_2 f_3)
\no \\ &&
- 4 (f_1 f_2 f_3 f_4)  - 4 (f_1 f_3 f_2 f_4)  - 4 (f_1 f_2 f_4 f_3)
\eea
using the notations,
\bea
f_i ^{\mu \nu} & \equiv & \e_i ^\mu k_i ^\nu - \e_i ^\nu k_i ^\mu
\no \\
(f_i f_j) & \equiv & f_i ^{\mu \nu } f_j ^{\nu \mu}
\no \\
(f_i f_j f_k f_l) & \equiv & f_i ^{\mu \nu } f_j ^{\nu \rho} f_k ^{\rho \sigma}
f_l ^{\sigma \mu}.
\eea
The expression $\R^4$, used in the preceding subsection, is related to $K$ 
in the following manner, $K \bar K = 2^6 \R^4$.
The moduli spaces $\M_1$ and $\M_2$ are described in Appendix A.
Finally, the various overall normalization constants   $C_0$ in , $Q_0$ in , $C_1$ 
and $C_2$ are given as follows,
\bea
\label{Cees}
C_0 = {1 \over 2^6}  \hskip .5in
& \hskip 1in &
Q_0 =  {\sqrt{2} \over 64 \pi^6 (\alpha ')^5 }
\no \\
C_1 = {1 \over 2^8 \pi ^2 (\alpha ')^5} & \qquad & 
C_2 = { \sqrt{2} \over 2^6  (\alpha ')^5} 
\eea
The constants $C_0$ and $Q_0$ for the tree-level amplitude will be calculated 
in Appendix B, $C_1$ will be calculated in Appendix C, and $C_2$ will
be calculated by factorization in section 3.

\subsection{Matching S-Duality and Perturbative predictions}

For tree-level, one- and two-loop, we have from (\ref{tree4}), (\ref{one4}),
and we shall establish in (\ref{D4R4}),
\bea
\label{perturbativeA}
\A_0 ^{(4)} (\e_i, k_i)
& = &
2 \pi C_0 \, Q_0 \,    e^{-2\lambda} \, \kappa ^4\, K \bar K \,
{\Gamma (-\alpha ' s/4 )\Gamma (- \alpha ' t/4)\Gamma (- \alpha ' u/4)
\over
\Gamma (1+\alpha ' s/4)\Gamma (1+\alpha ' t/4)\Gamma (1+\alpha ' u/4)}
\no \\
\A_1 ^{(R^4)} (\e_i, k_i)
& = &
{32 \pi \over 3} C_1  \, \kappa ^4 \, K \bar K \, 
\no \\
\A _2 ^{(D^4R^4)}  (\epsilon_i, k_i) 
& = &
{2 ^6 \pi ^3 \over 270}  C_2 \, e^{2 \lambda} 
\, \kappa ^4 \, K \bar K \, (\alpha ')^2 (s^2 + t^2 + u^2)
\eea
The one-loop term was approximated to order $R^4$,
using $\int _\Sigma d^2 z_i = 2 \, \Im \, \tau$, the 
formula (\ref{one4}) for the one-loop 4-point function,
and the fact that $\int _{\M_1}|d\tau|^2 (\Im \, \tau )^{-2}= 2 \pi / 3$.

\medskip

The predictions of S-duality and superstring perturbation theory
require the matching of (\ref{tree0}), (\ref{A1R4}), and (\ref{D4R41})
with the three expressions in (\ref{perturbativeA}).
Using the conversion relation $K \bar K = 2^6 \R^4$, these matching 
conditions are equivalent to the following relations,
\bea
(h=0) \hskip 1.1in
\kappa _{10}^2 e^{-2 \phi} 
& = & 
2 \pi \, C_0 Q_0 \kappa ^4 e^{- 2 \lambda}  \, 2^6
\no \\
(h=1) \hskip 1.1in
\kappa _{10}^2 \, {2 \pi ^2 \over 3} 
& = & 
{32 \pi \over 3} \, C_1  \, \kappa ^4 \, 2^6
\no \\
(h=2) \hskip .505in
\kappa _{10}^2 e^{2 \phi} { 4 \pi^4 \over 270} \, {(\alpha ')^2 \over 2^4}
& = & 
\rho { 2 ^6 \pi ^3 \over 270} \, C_2 \kappa ^4 e^{2 \lambda} (\alpha ')^2 \, 2^6
\eea
These three relations must hold for effectively two unknowns,
namely $\kappa _{10}^2 /\kappa ^4$ and $\exp \{ \phi - \lambda\}$.
Matching thus requires that a single relation between the 
coefficients $C_0 Q_0, \, C_1$ and $C_2$ hold,
\bea
\label{fact2}
C_1 ^2 = 2 \pi^2 C_0 Q_0 C_2
\eea
As will be shown in section 3, equation (\ref{factrel}), this relation 
is precisely the factorization condition on the two-loop 4-point function
used to determine $C_2$, and is manifestly satisfied by (\ref{Cees}).

\medskip

One also obtains the relation between the couplings 
$\phi$ and $\lambda$, using the specific values of the constants $C_0,Q_0,C_1$,,
\bea
e^{2 \phi} = 2^6 \sqrt{2} \,  \pi^2 e^{2 \lambda}
\eea
Notice that $\rho$ does not depend upon the details of the constants 
$C_0,C_1,C_2,Q_0$, as long as they satisfy the factorization equation 
(\ref{fact2}).

\newpage

\section{Normalization of superstring two-loop amplitudes}
\setcounter{equation}{0}

The expression for the two-loop 4-point function, derived in \cite{VI,zwzIII}, 
gives the amplitude only up to an overall constant factor $C_2$ which is  independent of moduli and of the Mandelstam variables. This constant $C_2$ is 
\`a priori unknown due to the fact that the chiral determinants of the Dirac operators on Riemann surfaces are themselves known only up to multiplicative constants depending only on the genus. In the present section we determine the exact value of $C_2$ from physical factorization to lower-loop amplitudes. 

\medskip

For convenience, we recall the two-loop 4-point function of (\ref{TypeII}),
\bea
\A_2 ^{(4)} (\epsilon_i, k_i)
=
C_2 \, e^{2 \lambda} \, K \bar K  \, \kappa ^4 \,
\int_{\M_2} { |d^3 \Omega|^2 \over (\det \Im \Omega)^5}
\int_{\Sigma^4}
|{\cal Y}_S|^2
{\rm exp}\bigg(-{\alpha'\over 2}\sum_{i<j}k_i\cdot k_j\,G(z_i,z_j)\bigg)
\eea
The antisymmetric biholomorphic 1-form $\Delta$ is defined 
in (\ref{Delta}), and  $\Y_S$  in (\ref{YS}).
To calculate $C_2$, we shall work out the separating degeneration 
limit of ${\cal A}_2^{(4)}$  to two one-loop amplitudes. Thus, we need the
precise asymptotics, including constants depending only on the genus, of the Green's function $G(z,w)$, of the quantity $\Y_s$, and
of the volume forms, in the  limit where the genus 2 surface 
$\Sigma$ degenerates to two separated genus one surfaces 
$\Sigma _1 \cup \Sigma _2$. 

\subsection{Degeneration formulas for $E(z,w)$, $G(z,w)$, and $\Y_s$}

We begin with the degeneration formulas for the period matrix, abelian differentials, and prime forms. They are all well-known \cite{ vv, dp88,fay}. However, in order to obtain precise values for the constants of importance to us, 
consistently with our notations, we provide a detailed derivation of the precise asymptotics for the prime form.  

\medskip

We shall right away restrict our considerations to genus 2. Choose a standard basis for the homology 1-cycles,
$A_I,B_I$, with $I=1,2$ and $\#(A_I,B_J)= \delta _{IJ}$.
The period matrix will be parametrized by
\bea
\Omega = \left ( \matrix{\tau_{11}' & \tau_{12} \cr \tau_{12}  & \tau _{22}' \cr}\right )
=
\left ( \matrix{\tau_{11} & 0 \cr 0 & \tau _{22} \cr}\right ) + \O(t)
\eea
The degeneration limit corresponds to letting $t \to 0$, so that $\tau_{12} \to 0$
and $\tau_{II}'$ tend to finite limits $\tau_{II} ' \to \tau _{II}$. The moduli $\tau_{11}$
and $\tau_{22}$  are then the moduli of the separated tori.
The resulting asymptotics may be expressed in terms of genus~1
$\tet$-functions. We recall the definition, mainly in order to fix our conventions
\bea
\tet  [\kappa _I ] (z _I ,\tau _{II})  \equiv  \sum _{m\in {\bf Z}}
\exp \biggl \{  i \pi (m+ \kappa _I ')^2 \tau  _{II}
+ 2 \pi i (m+\kappa _I ') (z_I + \kappa _I '') \biggr \}
\eea
The $\tet$-function satisfies the heat equation
$\label{heateq}
\p _{z_I} ^2 \tet  [\kappa _I] (z_I ,\tau _{II})
= 4 \pi i \p _{\tau_{II}}  \tet  [\kappa _I ](z_I , \tau_{II})$,
and the product relations $\tet '_1 (0,\tau_{II} ) = \tet [\nu_0]'(0,\tau_{II} ) =
 - \pi \ \tet  [\mu _2] \tet  [\mu _3]
\tet  [\mu _4] (\tau _{II}) = -2 \pi \eta (\tau_{II} )^3$,
where $\kappa_I$, $I=1,2$, stand for any genus 1 spin structures while the
spin structures $\mu_2$, $\mu_3$, $\mu_4$ are the three distinct even spin
structures, and $\nu_0$ is the unique odd spin structure. As usual, we 
set $\tet _1 (z,\tau) \equiv \tet [\nu_0] (z,\tau)$.
Formulas for $\tet$-function degenerations are standard,
and we have  (see e.g. \cite{IV}, eq. (5.10)),
\bea
\label{tetlimit}
\tet \left [ \matrix{\kappa _1 \cr \kappa _2 \cr} \right ] (z, \Omega )
=
\sum _{p=0} ^\infty {1 \over p!} \left ( { \tau_{12} \over 2 \pi i} \right )^p
\p^p _{z_1} \tet  [\kappa _1] (z_1, \tau_{11}')
\p^p _{z_2} \tet  [\kappa _2] (z_2, \tau_{22}')
\eea
where $z=(z_1,z_2) \in {\bf C}^2$.
In practice, we shall only need the following special cases,
\bea
\tet \left [ \matrix{\mu \cr \nu_0 \cr} \right ] (z, \Omega )
& = &
\tet  [\mu ](z_1, \tau_{11}) \, \tet _1 (z_2,\tau_{22}) + \O(t)
\no \\
\p_1 \tet \left [ \matrix{\mu \cr \nu_0 \cr} \right ] (0, \Omega )
& = &
2 \tau_{12} \p_{\tau_{11}} \tet  [\mu ](0, \tau_{11}) \, \tet _1  ' (0,\tau_{22}) 
+ \O(t^2)
\no \\
\p_2 \tet \left [ \matrix{\mu \cr \nu_0 \cr} \right ] (0, \Omega )
& = &
\tet  [\mu ](0, \tau_{11}) \, \tet _1 ' (0,\tau_{22}) + \O(t)
\eea
where $'$ denotes differentiation of the first argument, as usual, and $\mu$ 
is any even spin structure. 

\medskip

The degeneration formulas for the holomorphic Abelian differentials 
are given by \cite{fay} and \cite{dp88},
eq. (6.59). Their expressions for general $\Omega$
are denoted by $\o ^t _I (z)$, $I=1,2$, while the  holomorphic
differentials on the separated components  $\Sigma_I$ are $\v _I (z)$.
Also, we denote by $\v _p ^{(I)} (z)$ the Abelian differential of the second
kind on $\Sigma _I$ with a double pole at $z=p$, and unit residue.
(To be completely clear, if $z \in \Sigma _I$, then $\v _J (z)=0$ and
$\v^{(J)} _p (z) = 0$  when $J \not=I$.) The asymptotics to first order in $t$
is then given by
\bea
\o_1 ^t (z) & = & \left \{ \matrix{
\v _1 (z) + {t \over 4} \v_1 (p_1) \v_{p_1} ^{(1)} (z)
& z \in \Sigma _1 \cr
 {t \over 4} \v_1 (p_1) \v_{p_2} ^{(2)} (z)
& z \in \Sigma _2 \cr} \right .
\no \\ && \\
\o_2 ^t (z) & = & \left \{ \matrix{
 {t \over 4} \v_2 (p_2) \v_{p_1} ^{(1)} (z)
& z \in \Sigma _1 \cr
\v _2 (z) + {t \over 4} \v_2 (p_2) \v_{p_2} ^{(2)} (z)
& z \in \Sigma _2 \cr} \right .
\no
\eea
Here, the points $p_1$ and $p_2$ are the limiting points on the
surfaces $\Sigma _1$ and $\Sigma_2$ respectively, left of the
degeneration limit.
The normalizations of these differentials are as follows,
\bea
\int _{A_I} \v_J =\delta _{IJ}
& \hskip 1in &
\int _{B_I} \v_J = \delta _{IJ} \tau_{II}
\no \\
\int _{A_I} \v ^{(J)}_p   = 0
& \hskip 1in &
\int _{B_I} \v ^{(J)} _p  = 2 \pi i \delta _{IJ} \v _I (p)
\no \\
\int _{A_I} \o ^t _J =\delta _{IJ}
& \hskip 1in &
\int _{B_I} \o ^t _J = \Omega _{IJ}
\eea
It follows that we have\footnote{These formulas are naturally
interpreted as the period matrix deformations due to a Beltrami
differential supported at the points $p_1$ and $p_2$, with magnitude $t$.}
$\tau_{12}  = { \pi i \over 2} t \, \v_1 (p_1) \v_2(p_2) + \O(t^2)$,
$\tau_{11} '= \tau_{11} + { \pi i \over 2} t \, \v_1 (p_1) ^2 + \O(t^2)$,
$\tau_{22} ' = \tau_{22} + { \pi i \over 2} t \, \v_2 (p_2) ^2 + \O(t^2)$.
Parametrizing the tori $\Sigma_I$ by the standard parallelogram,
with vertices $0,1,\tau_{II} , \tau_{II} +1$, gives
$\v_I(z)=1$ for $z\in\Sigma_I$ and $v_I(z)=0$
for $z\notin\Sigma_I$.
Hence we have the following simplified expressions,
\bea
\tau_{12}={ \pi i \over 2} t  + \O(t^2),
\quad
\tau_{11} ' = \tau_{11} + { \pi i \over 2} t  + \O(t^2),
\quad
\tau_{22} '= \tau_{22} + { \pi i \over 2} t  + \O(t^2).
\eea

\medskip
We can derive now the asymptotics for
the genus 2 prime form $E(z,w)$. In general, $E(z,w)$ is defined by
\bea
E(z,w;\Omega) = { \tet [\nu ] \left ( \int ^z _w \o_I ^t , \Omega \right )
\over h_\nu (z,\Omega) h_\nu (w,\Omega)}
\eea
Here, $\nu$ is a genus 2 odd spin structure; $E(z,w;\Omega)$ is actually
independent of $\nu$, and we shall choose it as below.
The spinor $h_\nu$ is defined by,
\bea
h_\nu (z,\Omega)^2  =  \o_I^t (z) \p_I \tet [\nu ] (0, \Omega )
\hskip 1in
\nu =  \left [ \matrix{\mu \cr \nu_0 \cr} \right ]
\eea
We shall need the degeneration limit only when $z\in \Sigma _1$ and
$w \in \Sigma _2$, which we henceforth assume to be the case.
To leading order, we have
\bea
\int ^z _w \o_1 ^t & = & \int ^z _{p_1} \v _1 = z-p_1 + \O(t)
\no \\
\int ^z _w \o_2 ^t & = & \int ^{p_2}  _w \v _2 = -w  + p_2 + \O(t)
\eea
To leading order, the limit of the $\tet$-function factor in the prime form gives
\bea
\tet [\nu ] \left ( \int ^z _w \o_I ^t , \Omega \right )
=
\tet  [\mu ](z-p_1,\tau_{11} ) \, \tet _1  ( p_2-w, \tau_{22} ) + \O(t)
\eea
The calculation of the spinors $h_\nu$ proceeds analogously,
and we have
\bea
h _\nu (z, \Omega )^2 h_\nu (w,\Omega)^2
=
\tet  [\mu ](0,\tau_{11})^2 \tet _1 '(0,\tau_{22})^2
\left [
2 \tau \p_{\tau_{11}} \ln \tet  [\mu ] (0,\tau_{11}) + {1 \over 4 } t \v ^{(1)} _{p_1} (z)
\right ]
\eea
Using again the heat equation to recast the $\tau_{11}$-derivative in 
terms of $z$-derivatives, and the relation between $\tau_{12}$ and $t$,  
the term in brackets  becomes,
\bea
[\cdots ] =
 {t \over 4} \left \{
\p_z \p_{p_1} \ln \tet _1 (z-p_1, \tau_{11})
+
 \p_z^2 \tet  [\mu ](0,\tau_{11}) / \tet  [\mu ](0,\tau_{11}) \right \}
=
 { t \over 4} S_\mu (z,p_1)^2
\eea
where $S_\mu$ is the genus 1 Szeg\"o kernel for even spin structure $\mu$.
The first and second lines above are related by a form of the Fay identity
for genus 1. The Szeg\"o kernel is given by
\bea
S_\mu (z,p_1) =
{\tet [\mu](z-p_1,\tau_{11}) \tet _1 '(0,\tau_{11})
\over
\tet [\mu ] (0,\tau_{11}) \tet _1 (z-p_1,\tau_{11})}
\eea
Putting all together, we have
\bea
h_\nu (z) h_\nu (w)
=
\half \sqrt{t} \,
{\tet [\mu] (z-p_1,\tau_{11}) \tet _1' (0,\tau_{11}) \tet _1' (0,\tau_{22}) \over
\tet _1 (z-p_1, \tau_{11}) }
\eea
As a result, we have
\bea
E(z,w;\Omega ) & = &
2 t^{-\half} E_1 (z,p_1;\tau_{11} ) E_1 (p_2,w;\tau_{22} ) + \O(t^\half)
\no \\
& = &
(2\pi i /\tau_{12} )^\half  
E_1 (z,p_1;\tau_{11} ) E_1 (p_2,w;\tau_{22} ) + \O(\tau_{12}  ^\half)
\eea
where the genus 1 prime form is given by 
$ E_1 (z,w;\tau) = \tet _1 (z-w;\tau) / \tet _1 '(0,\tau)$.

\medskip

Next, we derive the asymptotics of
the Green's function $G(z,w)$.
The Green's function $G(z,w)$ was defined in (\ref{scalarprop}).
Its leading asymptotics arises from the degeneration of the prime form and produces a $\ln |t|$ term. The subleading asymptotics
is constant as $t \to 0$, and must also be retained. Asymptotics
in lower order terms are not needed. The precise formula is
quite simple,
\bea
G (z,w;\Omega)
& = &
- \ln |E(z,p_1;\tau_{11}) |^2 + 2 \pi (\Im \tau_{11})^{-1} \left ( \Im (z-p_1) \right )^2
\no \\ &&
- \ln |E(w,p_2;\tau_{22}) |^2  + 2 \pi (\Im \tau_{22} )^{-1} \left ( \Im (w-p_2) \right )^2
\no \\ &&
+ \ln (|t|/4) + o(1)
\eea
or, using the form of the genus 1 Green functions,
\bea
G (z,w;\Omega) =  \ln (|\tau_{12} |/ 2\pi ) +
G(z,p_1;\tau_{11}) + G(w,p_2;\tau_{22}) + o(1).
\eea

\medskip
Finally, making use of the limits of the holomorphic Abelian differentials
established earlier,
$\lim_{\tau\to 0}\o_I(z)=1$ or $0$ depending on whether
$z\in\Sigma_I$ or not, as well as
the fact that $s+t+u=0$,  we readily find the asymptotics for $\Y_S$,
\bea
\lim _{\tau \to 0} \Y_S = -s.
\eea

\subsection{Combinatorics of the degeneration}

Three important combinatorial considerations  need to be
taken into account in order to obtain the correct normalization 
of the 2-loop measure and amplitude from factorization.

\begin{enumerate}
\item 
The formulas for the Siegel fundamental domain require that
the genus 1 components of the degeneration be ordered \cite{siegel, moore}, 
in the sense that
\bea
\Im (\tau_{11}) \leq \Im (\tau_{22})
\eea
For an integrand that is symmetric under $\tau_{11} \leftrightarrow \tau_{22}$,
this ordering is equivalent to taking the  product over the two genus 
1 moduli spaces, and including a factor of 1/2.

\item
The pair $z_1,z_2$ must
run over both genus 1 components of the degeneration. 
This produces a factor of 2 when the formula is written 
with $z_1,z_2$ running only over one of the genus 1 components.

\item
As pointed out in \cite{moore}, one must impose the identification $t \sim -t$,
or equivalently $\tau_{12} \sim - \tau _{12}$. The explanation is as follows.
The genus 1 components each have a (conformal) automorphism $z \to -z$. Applying this transformation to only one genus 1 component changes 
$t \to -t$, in view of the plumbing fixture definition $ t =zw$, where 
$z$ and $w$ lie on opposite genus 1 components.

\end{enumerate}

\medskip

The two combinatorial factors due to points 1. and 2. above cancel 
one another if one takes the prescription that one integrates over the 
full genus 1 moduli  spaces of the two components and assumes that the pair 
$z_1,z_2$ runs only over a single component, while the other 
pair $z_3,z_4$ runs over the other component only.

\subsection{Factorization of the  two-loop 4-point amplitude}

We shall now evaluate the contribution to the residue of the pole in $s$ at  $s=- q^2$ where $q= k_1 + k_2 = -k_3 - k_4$ of the
two-loop amplitude resulting from the  separating degeneration alone. There are other singularities at $s= - q^2$, which are of no interest to us. The first is when two  vertex operators come close together, producing a pole;
the second is when a non-separating degeneration occurs,
producing a branch cut\footnote{For a detailed discussion
of the poles and branch cuts for the superstring box graph,
see \cite{dp92}.}. 

\medskip

By translation invariance on each torus, the points $p_1$ and $p_2$
are arbitrary, and the three genus 2 moduli $\tau_{11}$, $\tau_{22}$ and $\tau_{12}$ span the moduli coordinates on the two tori components 
$\Sigma _1$, $\Sigma_2$
as well as the plumbing fixture, without the need for $p_1$ and $p_2$.
Thus, to leading asymptotics, the measure has the following limit,
\bea
\lim _{\tau _{12} \to 0}  { |d^3 \Omega |^2 \over (\det \Im \Omega)^5}
= |d \tau_{12} |^2
{|d\tau _{11} |^2 \over (\Im \tau_{11})^5}
{|d\tau _{22}  |^2 \over (\Im \tau_{22})^5}
\eea
Combining all ingredients $G(z,w)$ and $\Y_S$ for the limit, as calculated above, we get
\bea
\A_2 ^{(4)} (\e_i, k_i) \sim
s^2  C_2 \, K \bar K \, (2 \pi )^{\alpha ' s/2} \, 
\int  |d \tau_{12} |^2 |\tau_{12}  | ^{-\alpha ' s/2}
\B_1 (k_1, k_ 2, q) \B_1(k_ 3, k_ 4,-q)
\eea
where the 3-point functions are given by (we suppress the dependence 
of $\B_1 ^{(3)}$ on the polarization vectors $\e_i$),
\bea
\label{bees}
\B_1 ^{(3)} (k_1, k_2,-q) & = &
\int _{\M_1} {|d\tau _{11} |^2 \over (\Im \tau_{11})^5}
\int \! d^2 \! z_1  d^2 \! z_2
\exp  {\alpha ' s \over 4} \left \{  G (z_1,z_2) -  G(z_1,p_1) -  G(z_2,p_1) \right \}
\no \\ && \\
\B_1 ^{(3)} (k_ 3, k_ 4,q) & = &
\int _{\M_1} {|d\tau _{22} |^2 \over (\Im \tau_{22})^5}
\int \! d^2 \! z_3   d^2 \! z_4
\exp   {\alpha ' s \over 4} \left \{   G (z_3,z_4) - G(z_3,p_2) -  G(z_4,p_2) \right \}
\no
\eea
The above kinematical factors are exactly as one would expect
for a 3-point function with the external momenta $(k_1,k_2,-q)$
in the first and $(k_3,k_4,q)$ in the second, in view of the 
kinematical relations
\bea
- k_1 \cdot (-q) = - k_2 \cdot (-q) =
- k_3 \cdot q = - k_4 \cdot q =  - \half s
\eea
Notice that, at the value $s=4/\alpha'$, the 3-point functions $\B_1 ^{(3)}$
themselves have a pole in $s$. This is as expected, and the pole arises
from the integration regions $z_1 \sim z_2$ and $z_3 \sim z_4$.
The factorization formula should be understood to hold for
$s$ in the neighborhood of $4/\alpha'$.

\medskip

It remains to evaluate the pole itself. We are interested only in the first
massive pole at $s=4/\alpha '$, which is also the first singularity (i.e. the 
smallest value of $s$ for which there is a singularity) of the
$\tau_{12} $-integral.
Both $\B_1$ factors are independent of $\tau_{12}$ in the limit 
$\tau_{12} \to 0$. Recall from the previous subsection that one must 
identify $\tau_{12} $ with $- \tau_{12}$, so that the integration over 
$\tau_{12}$ is actually over a disk with opposite identified, $D_\ep / \bZ_2$, 
where 
$D_\ep = \{ \tau_{12} \in \bC, ~ |\tau_{12} |<\ep\}$. Instead of the disk,
it is convenient to cutoff the $\tau_{12}$-integral by an exponential instead
with $\tau_{12} \in \bC$,
since the integral is then easily analytically continued in $s$. Retaining
only the pole parts, we have 
\bea
(2 \pi )^{\alpha ' s/2}  \int  |d \tau_{12} |^2 \, |\tau_{12} | ^{- \alpha ' s/2}
& = &
(2 \pi )^{\alpha ' s/2}  \int _{\bf C/\bZ_2}  |d \tau_{12} |^2  
|\tau_{12} | ^{-\alpha ' s/2 } e^{- |\tau|}
\no \\
& = &  -  { 16 \pi ^3 /\alpha ' \over s- 4/\alpha '}
\eea
Here the $\bZ_2$ factor accounts for the identification 
$\tau_{12} \to - \tau_{12}$.
Putting all together, we have
\bea
\A_2 ^{(4)} (\e_i, k_i)
=
- \delta (k) { 2^6  \pi ^3 C_2/\alpha ' \over s-4/\alpha' } \, 
e^{2 \lambda} \, K \bar K \,
\B_1 ^{(3)} (k_1,k_2,-q) \B_1 ^{(3)} (k_3,k_4,q).
\eea
It should be understood here that we are concerned only with that region of moduli space where the points on each torus are kept separated from one another.

\subsection{Factorization of the one-loop 4-point amplitude}

From the normalized one-loop 4-point amplitude $\A_1 ^{(4)}$ 
of (\ref{one4}), (derived in Appendix C), we obtain the 
residue at the massive pole  $s=4/\alpha'$, which arises when
the two insertion points $z_3$ and $z_4$ come close together.
In the region of moduli space where $z_ 4$ comes close to $z_3$, 
we eliminate $z_4$ in favor of $z= z_4-z_3$, but keep all other points. 
In this way, we get
\bea
\label{oneloopfactor}
\A_1 ^{(4)} (\e_i, k_i)
& = &
C_1  K \bar K \, \kappa ^4 \,
\int _{\M_1} {|d\tau|^2 \over (\Im \, \tau )^6} \int |dz|^2 \, |z|^{- \alpha ' s/2}
\prod _{i=1,2,3} \int d^2z_i
\no \\ && \hskip 1in \times
\exp \left \{ - {\alpha ' \over 2} \sum _{1 \leq i<j\leq 3} k_i \cdot k_j   G(z_i,z_j) \right \}.
\eea
The $z$-integral is familiar and, retaining only the pole parts in $s$, we have
\bea
\int |dz|^2 \, |z|^{- \alpha ' s/2} 
= 
\int _\bC |dz|^2 \, |z|^{- \alpha ' s/2} e^{-|z|}
= 
{ 8 \pi /\alpha '\over s - 4/\alpha'}.
\eea
The remaining integral is proportional to the 3-point function factors 
$\B_1$ introduced in the factorization of the 2-loop amplitude in (\ref{bees}). 
In (\ref{bees}), however, $\B_1$  was expressed as an integral over 2 of 
the three vertex  points, while the integral in (\ref{oneloopfactor})
has 3 vertex point integrations. Using translation invariance on the torus worldsheet, one of these three integrations may be carried 
out. This fixes the third point and produces a worldsheet volume 
factor $ 2 \Im \, \tau$. The factor of $\Im \, \tau $ reduces the denominator 
in (\ref{oneloopfactor}) from $(\Im \, \tau) ^{-6}$ to $(\Im \, \tau )^{-5}$ in
(\ref{bees}). The extra factor of 2 needs to be retained, and hence
\bea
\A_1 ^{(4)} (\e_i,k_i)
=
C_1  K \bar K \, \kappa ^4 \, { 16 \pi /\alpha '\over s - 4/\alpha'}
\B_1 (k_3,k_4,q), 
\hskip 1in q=-k_3-k_4.
\eea

\subsection{Factorization of the tree-level 4-point amplitude}

The normalized  tree-level 4-point  amplitude $\A_0 ^{(4)}$ of (\ref{tree4}),
(calculated in Appendix B),
may also be factored onto the  massive pole at $s=4/\alpha'$, and we find,
\bea
\A_0 ^{(4)} (\e_i,  k_i) =
{8 \pi \over \alpha '} C_0 \, Q_0 \, e^{-2\lambda}
 \kappa ^4 \, K \bar K \, { 1 \over s - 4/\alpha ' }.
\eea

\subsection{The factorization constraint}

Finally, we can implement the factorization constraint between the 
two-loop, one-loop, and tree-level superstring amplitudes.
Expressing the pole in terms of a common factor,
\bea
\A _0 ^{(4)} (k_i)
& = &
{ 8 \pi /\alpha '  \over s - 4/\alpha '} \, \kappa ^4 \, K \bar K
\bigg ( C_0 Q_0 e^{-2 \lambda} \bigg )
\no \\
\A _1 ^{(4)} (k_i)
& = &
{ 8 \pi /\alpha '  \over s - 4/\alpha '} \, \kappa ^4 \, K \bar K
\bigg ( 2 C_1 \B_1 ^{(3)} \bigg )
\no \\
\A _2 ^{(4)} (k_i)
& = &
 { 8 \pi /\alpha '  \over s - 4/\alpha '} \, \kappa ^4 \, K \bar K
\bigg ( 8 \pi^2  C_2 e^{+2 \lambda}  \B_1 ^{(3)} \B_1 ^{(3)}  \bigg
),\label{facmasss}
\eea
The particular factorization of the tree-level, one and two-loop string
amplitudes we are considering in (\ref{facmasss}) is illustrated in Figure 1.

\begin{figure}[tbph]
\begin{center}
\epsfxsize=4.0in
\epsfysize=4in
\epsffile{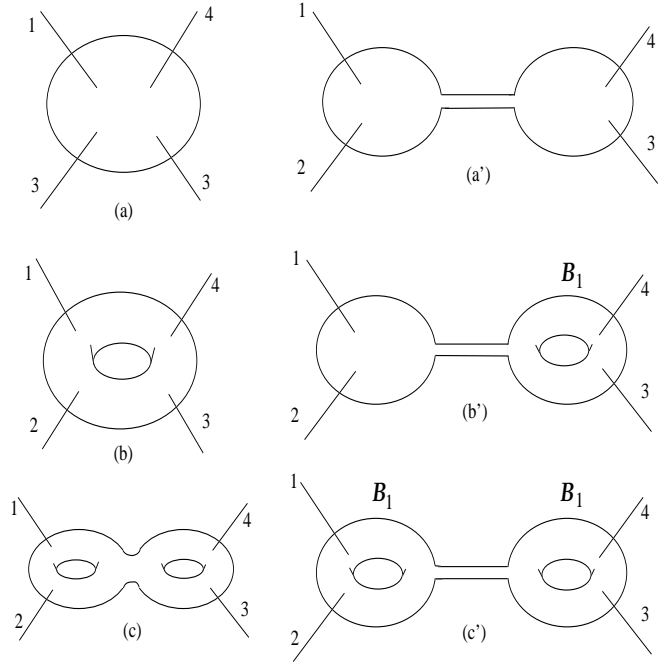}
\caption{Factorization of the tree-level (a) to (a'), one-loop (b) to (b') 
and two-loop (c) to (c')  4-point functions on a massive intermediate state.}
\label{fig}
\end{center}
\end{figure}

\bigskip

The consistency of the factorization of the two-loop amplitude on the
first massive pole with the tree-level and one-loop factorization
requires  the following relation 
\bea
\label{factrel}
C_1 ^2 = 2 \pi ^2 C_0 Q_0  C_2
\eea
Using the values obtained for the tree-level amplitudes in Appendix B
and for the one-loop amplitude in Appendix C,
\bea
C_0 = {1 \over 2^6} 
\hskip 1in 
Q_0 ={\sqrt{2} \over 2^6 \pi^6 (\alpha ')^5}
\hskip 1in
C_1 = {1 \over 2^8 \pi ^2 (\alpha ')^5}
\eea
we obtain
\bea
C_2 = { \sqrt{2} \over 2^6  (\alpha ')^5}.
\eea
The $\alpha'$ factor agrees on dimensional grounds.
For completeness, we list the two-loop 4-point function
expressed in terms of the S-duality normalization of the dilaton,
\bea
\A_2 ^{(4)} (\epsilon_i, k_i)
=
{e^{2 \phi} \, \kappa ^4 \, K \bar K   \over 2^{12} \pi ^2 (\alpha ')^5}  
\int_{\M_2} { |d^3 \Omega|^2 \over (\det \Im \Omega)^5}
\int_{\Sigma^4}
|{\cal Y}_S|^2
{\rm exp}\bigg(-{\alpha'\over 2}\sum_{i<j}k_i\cdot k_j\,G(z_i,z_j)\bigg)
\eea

\newpage

\section{The $D^4R^4$ term from the two-loop amplitude}
\setcounter{equation}{0}

In view of the presence of the factor $|\Y_S|^2$ in the two-loop
amplitude, and the fact that $\Y_S$ itself is linear in the kinematic
variables $s,t,u$, the term $D^4R^4$ will arise from setting $k_i=0$ in
the exponential in (\ref{TypeII}). The integration over $\Sigma ^4$
then reduces to the following integrals,
\bea
\int_{\Sigma^4} |{\cal Y}_S|^2
=
\int_{\Sigma^4} \bigg | s \Delta (1,4) \Delta (2,3) 
- t \Delta (1,2) \Delta (3,4) \bigg |^2
=
\L_1 + \L_2
\eea
where
\bea
\L_1
& = & (s^2 + t^2) \left ( \int _{\Sigma ^2} |\Delta (1,2)|^2 \right )^2
\no \\
\L _2 & = &
- 2 st \int _{\Sigma ^4} \Delta (1,4) \Delta (2,3)
\overline{\Delta (1,2)} \overline{\Delta (3,4)}
\eea
The integrations may be carried out using the Riemann bilinear relation,
\bea
\label{RR}
\int _\Sigma \omega _I  \omega _J ^* \equiv 
-i \int _\Sigma  \omega _I ^* \wedge  \omega _J= 2 \Im \Omega  _{IJ} 
\eea
The double integral for $\L_1$ and the quadruple integral for $\L_2$ are 
readily carried out using (\ref{RR}), and we find the following results,
\bea
\L_1 & = & 64 (s^2 + t^2) (\det \, \Im \, \Omega )^2
\no \\
\L_2 & = &    64 st \, (\det \, \Im \, \Omega )^2
\eea
Putting all together, and using $4 s^2 + 4 t^2 + 4st=2 (s^2 + t^2 + u^2)$, we have
\bea
\int_{\Sigma^4} |{\cal Y}_S|^2
=
32 (s^2 + t^2 + u^2)  (\det \, \Im \, \Omega )^2
\eea
In this limit, the amplitude reduces to the following expression,
\bea
\label{D4R40}
\A _2 ^{(D^4R^4)} (\epsilon_i, k_i)
=
8 V_2  C_2 \, e^{2 \lambda} (\alpha ')^2 (s^2 + t^2 + u^2)\, 
\kappa ^4  \, K \bar K  
\eea
Here, we  have restored the dependence on $\alpha'$ for later use,
and $V_2$ is the volume of genus~2 moduli space $\M_2$,
i.e. the volume of the fundamental domain of $Sp(4,\bZ)/{\bZ_2}$. Using the explicit formula for the volume given in Appendix \S A, we have 
\bea
\label{D4R4}
\A _2 ^{(D^4R4)} (\epsilon_i, k_i)
=
{2 ^6 \pi ^3 \over 270} 
 C_2 \, e^{2 \lambda} (\alpha ')^2 (s^2 + t^2 + u^2)\, \kappa ^4 \, K \bar K.
\eea

\newpage

\section{Heterotic $D^2F^2 F^2$ and $D^2 F^4$ terms}
\setcounter{equation}{0}

Extensive results are available on the Heterotic string contribution
to the low energy effective action  arising from 
tree-level and one-loop orders \cite{Cai:1986sa}. As the two-loop
contribution to the 4-point function is now also available, we
may add to these results as follows.
 
 \medskip
 
Two-loop 4-point amplitudes in the Heterotic string with space-time and gauge
structure $F^4$ and $F^2F^2$ were derived in \cite{VI} eq. (1.22),
and are given by
\bea
{\bf A} _{F^4}
=
C_H \bar K 
\int _{\M_2} { |d^3 \Omega|^2  \over (\det \Im \Omega )^5 \Psi _{10}(\Omega)}
\int _{\Sigma ^4} \W_{F^4} \bar \Y_S \exp \left ( - \sum _{i<j} k_i \cdot k_j
G(z_i,z_j) \right )
\eea
Here, $\bar K$ is the standard kinematical factor $2^3 \, t_8 F^4$,
$C_H$ is an overall normalization constant and $\Y_S$ is the same
quadri-holomorphic form familiar from the Type II amplitudes in (\ref{YS}).
The leading order term in small momenta is obtained when the exponential
factor equals 1 as $k_i \to 0$. The remaining factor $\bar \Y_S$ has two
extra factors of momenta. This guarantees that the $F^4$ and $F^2F^2$
terms receive no two-loop renormalization, as shown in \cite{VI}, 
but thus also implies non-vanishing  corrections of the type $D^2 F^4$
and $D^2 F^2 F^2$.

\medskip

We shall evaluate this contribution here for the simplest case,
namely for the $E_8 \times E_8$ theory,
when two external states (say 1,2) are in the first $E_8$,
while the remaining states (3,4) are in the second $E_8$.
The form of $\W_{F^4}$ is then particularly simple, and given by
\bea
\W_{F^4} = {1 \over 4} \tr (T^{a_1}T^{a_2}) \tr (T^{a_3}T^{a_4})
F_4 ^{(2)} (z_1,z_2) F_4 ^{(2)} (z_3,z_4)
\eea
In \cite{VI}, the function $F^{(2)}_4$ was computed explicitly in equation (12.7)
and is given by
\bea
F_4 ^{(2)} (z,w)
=
\Psi _4 \p_z \p_w \ln E(z,w) + {\pi i \over 2} \omega _I (z_1) \omega _J (z_2)
\p_{IJ} \Psi _4
\eea
Here, $\Psi _4 $ is the unique modular form of weight 4, defined by
$ \Psi _4 = \sum _\delta \tet [\delta ](0,\Omega )^8$.
Using the symmetries of $\Y_S$, the integration reduces to
\bea
\int _{\Sigma ^4} \W_{F^4} \bar \Y_S
=
-s \epsilon ^{IJ} \epsilon^{KL}
\int _{\Sigma ^4} F_4 ^{(2)} (1,2) F_4 ^{(2)} (3,4)
\omega _I ^* (1) \omega _J ^* (4)
\omega _K ^* (2)  \omega _L ^* (3)
\eea
The integral over $\Sigma ^4$ may be computed using (\ref{RR}) as
well as the Riemann relation
\bea
\int _\Sigma \omega ^*_I \, \p_z \p_w \ln E(z,w)
= 2 \pi  \omega _I (w)
\eea
It is convenient to first evaluate the double integral
\bea
\int _{\Sigma ^2}  F_4 ^{(2)} (1,2) \, \omega _I ^* (1) \omega _K ^* (2)
 =
4 \pi (\Im \Omega )_{IK} \Psi _4 + 2 \pi i   (\Im \Omega )_{IM} \p_{MN} \Psi_4
(\Im \Omega )_{NK}
\eea
and hence
\bea
\int _{\Sigma ^4} \W_{F^4} \bar \Y_S
=
 8 \pi ^2 s (\det \Im \Omega)^2 \, \Psi _4^2 \,
\det \bigg (\p_{IJ} \ln \Psi _4 - 2 i (\Im \Omega )^{-1} _{IJ} \bigg )
\eea
Putting all together, we have in this limit,
\bea
\A  _{F^4}
& = &
s \bar K \, 8 \pi ^2 C_H \, \tr (T^{a_1}T^{a_2}) \tr (T^{a_3}T^{a_4})
\int _{\M_2} {| d^3 \Omega|^2 \over (\det \Im \Omega )^3} \Psi_{F^4} (\Omega)
\no \\
\Psi _{F^4} (\Omega ) & = &
{\Psi _4(\Omega)^2 \over  \Psi _{10}(\Omega)}
\det \bigg [ \p_{IJ} \ln \left \{  \Psi _4 \, ( \det  \, \Im \, \Omega )^4 \right \} \bigg ]
\eea
where we have used the relation 
$ \p_{IJ} \ln \det \, \Im \, \Omega = -i/2 (\Im \, \Omega )^{-1} _{IJ}$.
Notice that $\Psi _{F^4}$ is a non-analytic modular function, i.e. a modular form
of weight 0. Indeed, combining the following  transformation laws,
\bea
\tilde \Omega & = & (A \Omega + B) (C \Omega +D)^{-1}
\no \\
\Psi _4 (\tilde \Omega ) & = &
\det (C \Omega + D)^4 \Psi _4 (\Omega)
\no \\
\det \, \Im \, \tilde \Omega & = & \left | \det (C \Omega + D) \right | ^{-2}  
\det \, \Im \, \Omega
\eea
we find the following transformation law,
\bea
\Psi _4 (\tilde \Omega)  (\det \, \Im \, \tilde \Omega )^4
=
\det (C \Omega ^* +D)^{-4} \, \Psi _4 (\Omega) \, (\det \, \Im \, \Omega )^4
\eea
Taking the log and differentiating in $\Omega _{IJ}$ produces a
rank 2 modular tensor
\bea
\p_{IJ} \ln \left  \{ \Psi _4 \, (\det \Im \Omega)^4 \right \}
\eea
whose determinant is a modular form of weight 2, making $\Psi _{F^4}$
a modular function.

\vskip .5in

\noindent{\Large \bf Acknowledgments}

\medskip

We are grateful to Costas Bachas, Michael Green, Sam Grushevsky, 
Boris Pioline, Jacob Sturm, Tomasz Taylor and Richard Wentworth 
for useful conversations,  correspondences, and references.

\newpage

\begin{appendix}

\section{The $Sp(2h,{\bf Z})/{\bf Z}_2$ fundamental domains}
\setcounter{equation}{0}

We give a synopsis of the key facts about genus 1 and 2 moduli
spaces and their measures needed in the sequel.

\subsection{Genus 1}

The genus 1 moduli space is given by the fundamental
domain of $Sp(2,\bZ)/\bZ_2$, namely
\bea
\M_1 = 
\left \{ \tau \in \bC; ~ \Im (\tau) >0, ~ |\tau|\geq 1, ~ |\Re (\tau)|\leq \half \right \}
\eea
Its volume for the Poincar\'e metric is as follows,
\bea
V_1 = \int _{\M_1} { |d\tau|^2 \over (\Im \, \tau )^2} = {2 \pi \over 3}
\hskip 1in 
|d\tau|^2 \equiv | d\bar \tau \wedge d\tau |
\eea

\subsection{Genus 2}

The genus 2 moduli space is considerably more complicated.
It is given by the fundamental domain of $Sp(4,\bZ)/\bZ_2$,
\bea
\M_2 = \left \{ \Omega = \left ( \matrix{\tau_{11} & \tau_{12} \cr \tau_{12} & \tau_{22} \cr } \right ) ~~ {\rm satisfying} ~ (1), \, (2), \, {\rm and} \, (3) \right \}
\eea
where the three conditions are given by, (see e.g. \cite{siegel} and \cite{moore}),
\bea
(1) & \hskip .5in & 
|\Re(\tau_{11})| \leq \half, ~ |\Re(\tau_{22})| \leq \half, ~ |\Re(\tau_{12})| \leq \half
\no \\
(2) & \hskip .5in & 
0 \leq |2 \Im (\tau_{12}) | \leq \Im (\tau_{11}) \leq \Im (\tau_{22})
\no \\
(3) & \hskip .5in & 
|\det (C\Omega + D)| \geq 1 ~~ {\rm for ~ all} ~~ 
\left ( \matrix{A & B \cr C & D \cr } \right ) \in Sp(4,\bZ)
\eea
The volume of the genus 2 moduli space was computed by Siegel
\cite{siegel}, and is given as follows,
\bea
\int _{\M_2} { d^3 \Re(\Omega) d^3 \Im (\Omega) \over (\det \, \Im \, \Omega )^3 } 
= { \pi ^3 \over 270}
\eea
In our conventions for the measure, this takes the form,
\bea
V_2 = \int _{\M_2} { |d^3 \Omega |^2 \over (\det \, \Im \, \Omega )^3 } 
= { 8 \pi ^3 \over 270}
\hskip 1in 
|d^3 \Omega |^2 = |\wedge _{I\leq J} d\tau_{IJ} |^2.
\eea
Notice that in the separating degeneration limit, $\M_2$ tends to the symmetrized product of $\M_1$ and $\M_2$, since their moduli
are ordered,
\bea
\lim _{\tau_{12} \to 0} \M_2 = \M_1 (\tau_{11}) \times \M_1 (\tau_{22})
\hskip .7in \Im (\tau_{11}) \leq \Im (\tau_{22})
\eea

\newpage

\section{Tree-level superstring amplitudes}
\setcounter{equation}{0}

In \cite{dp88}, an expression for the tree-level superstring amplitudes 
was derived only up to a constant due  to the combined determinants 
of matter and ghost fields on the sphere. The scalar as well as the $bc$ 
ghost determinants have been derived by Weisberger \cite{weis}. We give 
here a rigorous derivation for all spins. We also take the opportunity to 
correct some typos in \cite{dp88}. Since numerical constants are crucial 
for our present purposes, we explain their origin in some detail.

\subsection{Determinants of Laplacians on the sphere}

The metric on the round sphere with Gaussian curvature $R=1$
is given by,
\bea
\label{spheremetric}
g_{mn} = {2 \delta _{mn} \over (1 + |z|^2)^2}
\eea
where $\delta _{z\bar z}= \delta _{\bar z z}=1$ and 
$\delta _{zz}= \delta _{\bar z \bar z}=0$.
The Laplacians of interest are,
\bea
\Delta _{(n)} ^+ = -2 \nabla ^z _{(n+1)}\nabla _z ^{(n)}
\eea
for forms of spin $n\in \bZ/2$. The cases $n=0$, $n=1$, $n=-1/2$ and $n=+1/2$
respectively correspond to scalars, the $bc$ ghosts,
the Dirac spinor, and the $\beta \gamma$ superghosts.
Henceforth, we shall assume that $n\geq -1/2$;
in the contrary case, simply interchange $n$ and $-n-1$. 
The eigenvalues of the Laplacians are given by,
\bea
(\ell -n) (\ell + n +1)
 \qquad  \left \{ \matrix{ \ell =n+1 , n+2,  \cdots \cr
 {\rm multiplicity} ~ (2 \ell+1) \cr} \right .
\eea
Modes with $\ell =n$ and $l=-n-1$ correspond to zero modes
of $\nabla _z ^{(n)}$ and $\nabla ^z _{(n+1)}$, while modes with smaller
$\ell$  correspond to non-normalizable  solutions. The
determinant of the  Laplacian (with zero modes suitably removed),
is given by
\bea
\ln {\rm det} ' \Delta _n = - \zeta _{(n)} '(0)
\eea
where the corresponding $\zeta$-function is defined by
\bea
\zeta _{(n)} (s)
=
\sum _{\ell = n+1} ^\infty { 2 \ell + 1 \over (\ell -n)^s (\ell + n+1)^s}
=
\sum _{\ell = 1} ^\infty { 2 \ell + 2n+ 1 \over \ell ^s (\ell + 2n+1)^s}
\eea
Using Feynman parameters to combine both factors in the denominators of
the sum, and expressing the result with the help of the Hurwitz $\zeta$-function, 
which is defined by
\bea
\zeta (s,u) = \sum _{m=0}^\infty { 1 \over (m+u)^s}
\eea
we have
\bea
\zeta _{(n)} (s)
=
{\Gamma (2s-1) \over \Gamma (s) \Gamma (s-1)}
\int _0 ^1 d \alpha \, \alpha ^{s-2} (1-\alpha)^{s-2} \zeta (2s-1, 1+\alpha (2n+1))
\eea
The prefactor of $\Gamma$-functions vanishes to first order in $s$ at $s=0$, but the integral has poles at $s=0$. It suffices to isolate the parts
of the integrand that produce the poles. They arise from terms constant
and/or vanishing linearly at $\alpha =0,1$. We use the notation $\sigma = 2 s -1$,
and expand as follows,
\bea
\label{zetaexp}
\zeta (\sigma , 1+\alpha (2n+1)) 
& = &
\half \left \{ \zeta (\sigma , 1) + \zeta (\sigma , 2n+2) \right \}
+ a_1 (\alpha - \half)
+ \alpha (1 -\alpha) \varphi _n (\sigma ,\alpha)
\no \\ 
\varphi_n (\sigma , \alpha) & = &
\half \left \{ \varphi _n (\sigma ,0) + \varphi _n (\sigma , 1) \right \}
+ a_2 (\alpha - \half)
+ \alpha (1-\alpha) \psi _n (\sigma ,\alpha)
\qquad
\eea
The coefficients $a_1$ and $a_2$ are independent of $\alpha$, so that
the contribution of those terms to the integral over $\alpha$ vanishes
by the symmetry $\alpha \leftrightarrow (1-\alpha)$ of the integral.
The functions $\varphi_n$ and $\psi _n$ are analytic in $s$, just as $\zeta$
was. The integrals of the terms independent of $\alpha$ and linear in 
$\alpha(1-\alpha)$ are  calculated using the Euler beta-function integral.
The terms in $\varphi_n$ above may be computed by taking the derivative in $\alpha$  of (\ref{zetaexp}) at $\alpha =0,1$, and using the formula 
$\p  \zeta (\sigma, \alpha)/\p\alpha = - \sigma \zeta (\sigma +1, \alpha)$, 
\bea
\varphi_n (\sigma,1) + \varphi _n (\sigma,0)
=
\sigma (2n+1) \left ( \zeta (\sigma +1 ,2n+2) - \zeta (\sigma +1 ,1) \right )
\eea
Combining all results, we obtain,
\bea
\label{zetan}
\zeta _{(n)} (s) = \zeta _{(n)} ^0 (s) +
{\Gamma (2s-1) \over \Gamma (s) \Gamma (s-1)}
\int _0 ^1 d \alpha \, \alpha ^s (1-\alpha)^s \psi_n (2s-1, \alpha )
\eea
where the reduced function  is given by
\bea
\label{zeta0}
\zeta _{(n)} ^0 (s)
& = &
\zeta (2s-1, 1) + \zeta (2s-1, 2n+2) + \half (s-1)(2n+1)
\left ( \zeta (2s ,2n+2) - \zeta (2s,1) \right )
\no \\
& = & 
2 \zeta _R (2s-1) - \sum _{\ell =1} ^{2n+1} { 1 \over \ell ^{2s-1} }
+ \half (1-s) (2n+1) \sum _{\ell =1} ^{2n+1} {1 \over \ell ^{2s} }
\eea
Here $\zeta _R$ is the Riemann $\zeta$-function. The $\zeta ^0 _{(n)}$ 
part of (\ref{zetan}) admits an analytic continuation in $s$ throughout the 
complex plane $\bC$ because $\zeta _R$ does, while the integral term
in (\ref{zetan}) is analytic in $s$ as long as $\Re (s) > -1$. In particular, 
the entire expression is now well-defined and analytic around $s=0$,
and its value there may be readily computed.

\medskip

To evaluate $\zeta _{(n)} '(0)$, it  suffices to differentiate 
(\ref{zetan}) is $s$ and to set $s=0$,
\bea
\zeta _{(n)} '(0) = \zeta _{(n)} ^{0\, \prime} (0)
+ \half \int _0 ^1 d \alpha \, \psi _n (-1,\alpha)
\eea
To calculate $\psi _n (-1,\alpha)$, we use its definition in (\ref{zetaexp}),
combined with the following formula, which may be found in \cite{bateman},
\bea
\zeta (-1, u) =  \half u (1-u) - {1 \over 12}
\eea
As a result, $\varphi_n (-1,\alpha) = (2n+1)^2/2$ and thus
$\psi _n (-1,\alpha)=0$, so that the integral in (\ref{zetan}) vanishes.
The remaining terms are  obtained from differentiating (\ref{zeta0}),
and we find,
\bea
\zeta _{(n)} '(0) = 4 \zeta _R '(-1) - \half (2n+1)^2 + \sum _{\ell =1}^{2n+1}
(2\ell - 2n -1) \ln \ell
\eea
By the same methods, we may read off also the following useful values,
\bea
\zeta _{(n)} (0) = -n - {2 \over 3}
\eea
The special cases $n=0$, $n=\pm 1/2$, and $n=1$ give the following
$\zeta _{(n)}'(0)$,
\bea
\zeta _{(0)} '(0) & = & 4 \zeta _R '(-1) - \half
\no \\
\zeta _{-1/2} '(0) & = & 4 \zeta _R '(-1)
\no \\
\zeta _{+1/2} '(0) & = & 4 \zeta _R '(-1) -2 + 2 \ln 2
\no \\
\zeta _{(1)} '(0) & = & 4 \zeta_R '(-1) - {9 \over 2} + 3 \ln 3 + \ln 2
\eea
Finally, we use the general formula that 
$ \zeta _{\lambda \Delta _{(n)}} (s) = \lambda ^{-s} \zeta _{\Delta _{(n)}}(s)$
to rescale the Laplace operators and the determinants,
\bea
{\rm det} ' \lambda \Delta _{(n)} = \lambda ^{\zeta _{(n)} (0)} {\rm det} ' \Delta _{(n)}
= \lambda ^{-n-2/3} {\rm det} ' \Delta _{(n)}
\eea
Putting together all the determinants
that enter the superstring, we have
\bea
\left ( {4 \pi ^2 \alpha ' {\rm det} ' \Delta _{(0)}  
\over \int _\Sigma \sqrt{g} } \right )^{-5} \,
{\rm det} ' (2 \Delta _{(1)}) \, 
{ \left (  {\rm det} \Delta _{(-1/2)} \right )^5 \over {\rm det} ' (2 \Delta _{(1/2)}) }
=
{\sqrt{2} \over 27 (\pi \alpha ')^5}
\eea

\subsection{The conformal Killing vector volume}

The measure $|\delta \rho_b |^2$  of $SL(2,{\bf C})$  (denoted
$d\mu = |\delta \rho_b |^2$ in \cite{dp88}, eq.(2.108)), may be defined in a
variety of  related ways. It is convenient to express these in terms
of the analytic  measure $\delta  \rho_b $. First, we have the
formula of \cite{dp88}, eq.(2.108),
\bea
\label{drho1}
\delta \rho_b =
{\delta  z_0 \, \delta  z_1 \, \delta  z_\infty \over (z_0-z_1)  (z_1-z_\infty)
 (z_\infty -z_0) }
\eea
Parametrizing the points as the images under $g \in SL(2, {\bf C})$
of three fixed  points 0, 1, $\infty$,
\bea
g = \left ( \matrix{a  & b  \cr c  & d  \cr} \right )
\hskip 1in
z_n = {a  n + b  \over c  n + d }
\eea
for $n=0,1,\infty$ and $ad-bc =1$,
the measure $\delta \rho_b $ may be expressed in
terms of $a,b,c,d$,
\bea
\label{drho2}
\delta \rho_b  & = & 
- a  \, \delta b \wedge \delta c \wedge \delta d
+ b  \, \delta a \wedge \delta c \wedge \delta d
- c \, \delta a \wedge \delta b \wedge \delta d
+ d \, \delta a  \wedge \delta b \wedge \delta c
\no \\
& = & ( \delta a - \delta d ) \wedge \delta b  \wedge \delta c
\qquad {\rm at} ~ g=I
\eea
The calculation of the conformal Killing volume now proceeds as follows.
The integration over all vector fields $v^m$ parametrizing infinitesimal
metric deformations is only over those $v^m$ orthogonal to the
conformal Killing vector fields. On the other hand, we divide out
by the entire volume of all vector fields. We want to trade in the
measure $|\delta  \rho_b |^2$ for the integration over the conformal
Killing vector fields $v^m$. To see the effect, it suffices to determine
the ratio of these two measures. Given that we are dealing with
the invariant measure on a group $SL(2,{\bf C})$, we need to
do this just around the identity in $SL(2,{\bf C})$.

\medskip

The vector fields $v^z$ are the variations in $z$
due to variations in $a,b,c,d$,
\bea
\delta  v^z = \sum _{n=0,1,2} \delta  \epsilon _n z^n =
\delta  b + (\delta a  -  \delta d) z - \delta c z^2
\eea
Hence the volume form may be expressed in terms of the $\delta \epsilon_n$,
\bea
\delta  \rho_b
= - \delta \epsilon _0 \wedge \delta \epsilon _1 \wedge \delta \epsilon _2
\eea
On the other hand, the measure for the conformal Killing vectors 
$|\delta ^3v|^2$ is governed by the metric on the space of Killing vectors,
\bea
||\delta  v ||^2 = \int d^2 z \sqrt{g} g_{mn} \delta  v ^m \delta  v^n
=  \sum _{n=0,1,2} |\delta  \epsilon _n|^2 N_n^2
\eea
Hence we have $|\delta ^3 v|^2 = N_0 ^2 N_1^2 N_2^2 |\delta  \rho _b|^2 $. 
With the round metric of (\ref{spheremetric}),
the normalization factors are evaluated as follows,
\bea
N_n ^2 = 2 \int d^2 z (g_{z\bar z})^2 |z^n|^2
=
16 \pi \int _0 ^\infty dr^2 { r^{2n} \over (1+r^2)^4}
\eea
and we find, $N_0 ^2 =  N_1^2/2 = N_2^2 = 16 \pi/3$, 
in agreement with \cite{weis}, and
\bea
\label{CKVvols}
|\delta  \rho_b |^2 =
{1 \over N_0^2 N_1^2 N_2^2} |\delta ^3v|^2 =
2 \left ( {3 \over 16 \pi} \right )^3 |\delta ^3v|^2
\eea

\subsection{The superconformal Killing spinor volume}

The superconformal case is analogous to the conformal case, but
the group of interest is now the supergroup $OSP(1,1)$. The derivation
given here is parallel to the procedure of \cite{dp88} \S III. L, page 976. 
We isolate the measure $ |\delta \rho_s|^2$ on $OSP(1,1)$ 
in terms of $z_{ij} = z_i - z_j - \theta _i \theta _j$,
\bea
\label{drho3}
\delta \rho_s =
{\delta  z_0 \, \delta z_1 \, \delta  z_\infty \, \delta \theta _0 \,
\delta \theta _1 \, \delta  \theta _\infty
\over z_{01} \, z_{1\infty} \, z_{\infty 0}  }
(z_{0 1} \theta _\infty + z_{1\infty} \theta _0 + z_{\infty 0} \theta _1 + \theta _0 \theta _1 \theta _\infty)
\eea
The last factor is related to the invariant $\Delta$ of \cite{dp88,Aoki:1987yc}.
The $OSP(1,1)$ transformations act by
\bea
T(z, \theta) = \left ( {a z + b + \alpha \theta \over c z + d + \beta \theta}, ~
{ \gamma z + \epsilon + A \theta \over  c z + d + \beta \theta} \right )
\eea
where the matrices
\bea
T = \left ( \matrix{    a & b & \alpha \cr
                c & d & \beta \cr
                \gamma & \epsilon & A\cr} \right )
\hskip 1in
K = \left ( \matrix{    0  & +1  & 0 \cr
                -1 & 0    & 0 \cr
                0  & 0 & -1   \cr} \right )
\eea
satisfy $T^t K T=K$. We parametrize the variation of three points
\bea
(z_0,\theta _0) & = &  T (0,\theta)
\no \\
(z_1 , \theta _1 ) & = & T (1,0)
\no \\
(z_\infty , \theta _\infty) & = & T (z,0)
\eea
infinitesimally around the identity transformation $T=I$. The general
variation is given by
\bea
\delta z_n & = & \delta b + (\delta a - \delta d) z_n - \delta c z_n^2
+ (\delta \alpha  - z_n \delta \beta ) \theta _n
\no \\
\delta \theta _n & = &
\delta \epsilon + z_n \delta \gamma  + \delta A \theta _n
- \theta _n (z_n \delta c  + \delta d) + \delta \theta \delta _{n,0}
\eea
The measure may now be computed, and we find,
\bea
\delta \rho _s = \theta \, \delta \theta \wedge \delta \gamma \wedge \delta \epsilon
\wedge (\delta a - \delta d) \wedge \delta b \wedge \delta c
\eea
The normalizations for the two conformal Killing spinors are now given by
\bea
N_n ^2 = 2 \int d^2z (g_{z\bar z})^{3/2} |z^{n-\half } |^2
\qquad \qquad n = \half, \, {3\over 2}
\eea
Explicit evaluation gives
\bea
N_\half ^2 = N_{3 \over 2}  ^2 = 2^{5/2} \pi
\eea
The measures are related by
\bea
|\delta \rho _s |^2 = N_\half ^2 \, N_{3\over 2}  ^2 |\delta ^2 \zeta |^2 | \delta ^3 \rho_b|^2 
=
{ 27 \over 64 \pi}  |\delta ^3 v \, \delta ^2 \zeta |^2
\eea
The relation of $|\delta ^3 \rho_b|^2$ to $|\delta ^3 v|^2$ is as 
in the bosonic case (\ref{CKVvols}).

\subsection{The coefficient $Q_0$}

Following the gauge-fixing procedure of \cite{dp88}, the coefficient $Q_0$ is found to be given by
\bea
Q_0
=
{N_\half ^2 \, N_{3 \over 2} ^2 \over N_0^2N_1^2N_2^2}
\left ( {{\rm det}\,\Delta_{(-1/2)}
\over
\pi \alpha'{\rm det}' \Delta_{(0)}} \right )^5
{{\rm det}'(2\Delta_{(1)})
\over
{\rm det}' (2\Delta_{(1/2)})}
=
{\sqrt 2\over 64\pi^6(\alpha')^5}
\eea

\subsection{The massless tree-level $N$-point function}

This calculation was carried out in \cite{dp88}, III.L,
but some of the coefficients are inaccurate. The
correct expressions are as follows. Let
the superfield propagator on the superplane (projected from the supersphere) be given by,
\bea
\bG (\z , \z') = - \ln \left ( | z-z' - \theta \theta ' |^2 + \epsilon ^2 \right )
\eea
where $\epsilon$ gives the correct short-distance prescription.

\medskip

To compute the scattering amplitudes, it is actually convenient to
view $\epsilon$ and $\bar \epsilon$ as Grassmann odd numbers 
and to work with the vertex generating function
\bea
V^* (\epsilon, \bar \epsilon, k) 
= \int _\Sigma d^{2|2} \z \, E \, \exp \{ i k \cdot X
+ \epsilon ^\mu \D_+ X^\mu + \bar \epsilon ^\mu \D_- X^\mu \}
\eea
where it is understood that this quantity must be expanded precisely
to first order in $\epsilon $ and to first order in $\bar \epsilon$, 
a prescription that will be indicated by $|_{\e\bar \e}$.
We then have,
\bea
\< \prod _{i=1}^N V (\epsilon_i, \bar \epsilon _i, k_i) \>
=
\< \prod _{i=1}^N V ^* (\epsilon_i, \bar \epsilon _i, k_i) \> \bigg |_{\e \bar \e}
\eea
It is straightforward to compute now,
\bea
\< \prod _{i=1}^N V ^* (\epsilon_i, \bar \epsilon _i, k_i) \>
=
\prod _{i=1}^N \int _\Sigma d^{2|2}\z_i
\exp \left \{ \G_N - \sum _{i <j} k_i \cdot k_j \bG (\z_i,\z_j) \right \}
\eea
where
\bea
\G _N & = &
\sum _{i \not= j} ^N \bigg (
i k_i \cdot \e_j \D^j _+ + i k_i \cdot \bar \e _j \D_- ^j 
- \half \e_i \cdot \e_j \D_+ ^i \D_+^j 
- \half \e_i \cdot \bar \e_j \D_+ ^i \D_-^j 
\no \\ && \hskip 1in 
- \half \bar \e_i \cdot \e_j \D_- ^i \D_+^j 
- \half \bar \e_i \cdot \bar \e_j \D_- ^i \D_-^j  \bigg ) \bG (\z_i, \z_j)
\eea
Neglecting contact terms, as we are instructed to do by the ``cancelled 
propagator argument",  we have the following expressions,
\bea
\D_+ ^j \bG (\z_i,\z_j ) = - {\theta _{ij} \over z_{ij}}
& \qquad & \hskip .22in
\D_- ^j \bG (\z_i,\z_j ) = - {\bar \theta _{ij} \over \bar z_{ij}}
\no \\
\D_+^i \D_+ ^j \bG (\z_i,\z_j ) = - {1\over z_{ij}}
& \qquad & 
\D_-^i \D_- ^j \bG (\z_i,\z_j ) = - {1\over \bar z_{ij}}
\no \\
\D_+^i \D_- ^j \bG (\z_i,\z_j ) = 0 \hskip .26in
& \qquad & 
\D_-^i \D_+ ^j \bG (\z_i,\z_j ) = 0
\eea
where we use the notation,
\bea
\theta _{ij} = \theta _i - \theta _j 
& \qquad &
z_{ij} = z_i -z_j -\theta _i \theta _j 
\no \\
\bar \theta _{ij} = \bar \theta _i - \bar \theta _j 
& \qquad &
\bar z_{ij} = \bar z_i - \bar z_j - \bar \theta _i \bar \theta _j 
\eea
Thus, we have the following systematic expression for 
$\G_N= \G_N ^{(+)} + \G_N ^{(-)}$,
\bea
\G_N ^{(+)} & = &
\sum _{i\not= j} ^N \left (
\half \e_i \cdot \e_j {1 \over z_{ij}}  - i k_i \cdot \e_j { \theta _{ij} \over z_{ij} }  \right )
\no \\
\G_N ^{(-)} & = &
\sum _{i\not= j} ^N \left (
 \half \bar \e_i \cdot \bar \e_j {1 \over \bar z_{ij}}
- i k_i \cdot \bar \e_j { \bar \theta _{ij} \over \bar z_{ij} } \right )
\eea
The superconformal volume is factored out as follows,
\bea
\< \prod _{i=1}^N V ^* (\epsilon_i, \bar \epsilon _i, k_i) \>
& = & 
\left ( \int |\delta \rho _s|^2 \right ) \int d^2 \theta _{N-2}
\prod _{i=1}^{N-3}  \int  d^{2|2}\z_i  \W^{(+)} _N \W^{(-)} _N
\prod _{ i < j} ^ {N-1}  |z_{ij}|^{2 k_i \cdot k_j}
\no \\
\W _N ^{(+)} & = &
\lim _{z_N \to \infty }  \left ( z_N \exp \{ \G_N ^{(+)} \} 
\bigg |_{\e} \right )
\no \\
\W _N ^{(-)} & = &
\lim _{\bar z_N \to \infty}  \left ( \bar z_N \exp \{ \G_N ^{(-)} \} 
\bigg |_{\bar \e} \right )
\eea
Here, it is understood that 
\bea
z_{N-2} = \bar z_{N-2} =0
& \qquad &  \theta _N = \bar \theta _N=0
\no \\ 
z_{N-1} = \bar z_{N-1}=1
& \qquad & \theta _{N-1} = \bar \theta _{N-1}=0
\eea
and the subscript $|_\e$ (respectively $|_{\bar \e}$) stands for the  prescription of retaining only terms that are linear in each $\e_i$ (respectively  in each $\bar \e_i$).

\medskip

For $N=3$, only the integration over $\theta _1$ remains, and we have 
\bea
\W _3 ^{(+)} = 
i (\e_1 \cdot \e_2) (\e_3 \cdot k_1) +
i (\e_2 \cdot \e_3) (\e_1 \cdot k_2) +
i (\e_3 \cdot \e_1) (\e_2 \cdot k_3)  
\eea

\subsubsection{The 4-point function}

We expand $\G_4 ^{(+)} $ and use the fact that $\theta _3 = \theta _4=0$
to simplify the result. We obtain, 
\bea
\G_4 ^{(+)}  & = &
\e_1 \cdot \e_2 { 1 \over z_{12}} +
\e_1 \cdot \e_3 { 1 \over z_{13}} +
\e_1 \cdot \e_4 { 1 \over z_{14}} +
\e_2 \cdot \e_3 { 1 \over z_{23}} +
\e_2 \cdot \e_4 { 1 \over z_{24}} +
\e_3 \cdot \e_4 { 1 \over z_{34}} 
\no \\ && 
- i (k_1 \cdot \e_2 + k_2 \cdot \e_1) { \theta _{12} \over z_{12}}
- i (k_1 \cdot \e_3 + k_3 \cdot \e_1) { \theta _1 \over z_{13}}
- i (k_1 \cdot \e_4 + k_4 \cdot \e_1) { \theta _1 \over z_{14}}
\no \\ &&
- i (k_2 \cdot \e_3 + k_3 \cdot \e_2) { \theta _2 \over z_{23}}
- i (k_2 \cdot \e_4 + k_4 \cdot \e_2) { \theta _2 \over z_{24}}
\eea
Next, exponentiate while retaining only contributions that are 
linear in each $\e_i$,
\bea
\W_4 ^{(+)} 
& = &
- (\e_1 \cdot \e_2) (\e_3 \cdot \e_4) { 1 \over z_{12} } 
- (\e_1 \cdot \e_3) (\e_2 \cdot \e_4) { 1 \over z_{13} } 
- (\e_1 \cdot \e_4) (\e_2 \cdot \e_3) { 1 \over z_{23} } 
\no \\ &&
- 
(\e_1 \cdot \e_2) \left ( (\e_3 \cdot k_1) ( \e_4 \cdot k_2) 
{ \theta _1 \theta _2 \over z_{12} z_{13} }
+ (\e_4 \cdot k_1) ( \e_3 \cdot k_2) 
{ \theta _1 \theta _2 \over z_{12} z_{23} } \right ) 
\no \\ && + {\rm 5 ~ permutations~of~the~last~line}
\eea
The basic integral formula needed for 4-point functions is\footnote{We use
the following conventions (which differ from those used in \cite{dp88}),  
$d^2 z = -i d\bar z \wedge dz$, 
$d^2 \theta = d \bar \theta d \theta $, so that 
$\int d^2 \theta \, \theta \bar \theta =1$, and
$\int d^2 \theta _1 \int d^2 \theta _2 \, (\theta _1 \theta _2) 
(\bar \theta _1 \bar \theta _2) = -1$.}
\bea
\int _\bC d^2z \, z^A (1-z)^B \bar z^{\tilde A} (1 - \bar z)^{\tilde B}
=
2 \pi { \Gamma (1+A) \Gamma (1+B) \over \Gamma (2 + A + B)}
\cdot
{\Gamma (-1 - \tilde A - \tilde B) \over \Gamma (- \tilde A) \Gamma (- \tilde B)}
\eea
for $A - \tilde A, \, B - \tilde B \in \bZ$. 
The integral is invariant under  $(A,B) \leftrightarrow (\tilde A, \tilde B)$, 
and factors according to the left-moving $z$-dependence  of the exponents 
$A,B$, and the right-moving $\bar z$-dependence of the exponents 
$\tilde A, \tilde B$. The superspace integrals we need for $c,\tilde c =0,1$ 
are as  follows,
\bea
&&
\int d^{2|2} \z_1 \int d^2 \theta _2 
(\theta _1 \theta _2)^c (\bar \theta _1 \bar \theta _2)^{\tilde c}   \, 
z_{12} ^A (1-z_1)^B \bar z_{12} ^{\tilde A} (1 - \bar z_1)^{\tilde B}
\no \\ && \hskip 1in
=
2 \pi (-)^c { \Gamma (1+A) \Gamma (1+B) \over \Gamma (1  + A + B +c )}
\cdot
{\Gamma (-\tilde c - \tilde A - \tilde B) \over \Gamma (- \tilde A) \Gamma (- \tilde B)}
\eea
which is also invariant under the interchange 
$(A,B,c) \leftrightarrow (\tilde A, \tilde B, \tilde c)$.
The integrals needed here are given as follows, for
$a,\tilde a, b,\tilde b, c, \tilde c \in \bZ$,
\bea
&&
\int d^{2|2} \z_1 \int d^2 \theta _2 
(\theta _1 \theta _2)^c (\bar \theta _1 \bar \theta _2)^{\tilde c}   \, 
z_{12} ^{- {s \over 2} -a}  (1-z_1)^{- {u \over 2} -b} 
\bar z_{12} ^{ - {s \over 2} - \tilde a } (1 - \bar z_1)^{- {u \over 2} - \tilde b}
\no \\ && \hskip 1in
=
2 \pi \R \tilde \R { \Gamma (-s/2) \Gamma (-t/2) \Gamma (-u/2)
\over 
\Gamma (1+s/2) \Gamma (1+t/2) \Gamma (1+u/2)}
\eea
where 
\bea
\R & = & 
(-)^c {\Gamma (1-s/2-a) \Gamma (1+t/2) \Gamma (1-u/2-b)
\over 
\Gamma (-s/2) \Gamma (1+t/2 -a-b+c) \Gamma (-u/2)}
\no \\
\tilde \R & = &
{ \Gamma (1+ s/2) \Gamma (- t/2 +\tilde a + \tilde b - \tilde c) \Gamma (1+u/2)
\over
\Gamma (s/2 + \tilde a) \Gamma (-t/2) \Gamma (u/2 + \tilde b)}
\eea
The various values needed here for the calculation are given in the table below.

\begin{table}[htdp]
\begin{center}
\begin{tabular}{|c||c|c|c||c|c|c|} \hline 
form of prefactor & $a$ & $b$ & $c$ & $\R$ & sign & net factor \\ \hline \hline
$-1 / (z_{12} ) $ & 1 & 0 & 0 & $ - tu /4 $ & $-$ & $ + tu/4$
\\ \hline
$-1 / (z_{13} ) $ & 0 & 1 & 0 & $ - st /4 $ & $+$ & $ - st /4 $
\\ \hline
$-1 / (z_{23}) $ & 0 & 0 & 0 & $ + su /4 $ & $ +$ & $ + su /4 $
\\ \hline
$- \theta _1 \theta _2 / ( z_{12} z_{13}) $ & 1 & 1 & 1 & $ - t /2 $ & $+$ & $-t/2$
\\ \hline
$- \theta _1 \theta _2 / ( z_{12} z_{23}) $ & 1 & 0 & 1 & $+u/2$ & $+$& $+u/2 $
\\ \hline
$- \theta _1 \theta _2 / ( z_{13} z_{23}) $ & 0 & 1 & 1 & $+s/2$ & $+$& $+s/2 $
\\ \hline \hline
\end{tabular}
\end{center}
\end{table}%

\noindent
A similar table holds true for $\tilde \R$.

\subsubsection{Final formula for the 4-point function}

\bea
\W_4 ^{(+)} 
& \to &
+ {1 \over 4} t u (\e_1 \cdot \e_2) (\e_3 \cdot \e_4) 
- {1 \over 4} st (\e_1 \cdot \e_3) (\e_2 \cdot \e_4)  
+ {1 \over 4 } su (\e_1 \cdot \e_4) (\e_2 \cdot \e_3)  
\no \\ &&
-  \half t  (\e_1 \cdot \e_2)  (\e_3 \cdot k_1) (\e_4 \cdot k_2) 
- \half u (\e_1 \cdot \e_2)  ( \e_3 \cdot k_2) (\e_4 \cdot k_1) 
\no \\ && 
+ \half s (\e_1 \cdot \e_3)  (\e_2 \cdot k_3) ( \e_4 \cdot k_1)  
+  \half t  (\e_1 \cdot \e_3)  (\e_2 \cdot k_1) (\e_4 \cdot k_3) 
\no \\ && 
- \half s (\e_1 \cdot \e_4)  (\e_2 \cdot k_4) ( \e_3 \cdot k_1)  
-  \half u  (\e_1 \cdot \e_4)  (\e_2 \cdot k_1) (\e_3 \cdot k_4) 
\no \\ && 
- \half s (\e_2 \cdot \e_3)  (\e_1 \cdot k_3) ( \e_4 \cdot k_2)  
-  \half u  (\e_2 \cdot \e_3)  (\e_1 \cdot k_2) (\e_4 \cdot k_3) 
\no \\ && 
+ \half s (\e_2 \cdot \e_4)  (\e_1 \cdot k_4) ( \e_3 \cdot k_2)  
+  \half t  (\e_2 \cdot \e_4)  (\e_1 \cdot k_2) (\e_3 \cdot k_4) 
\no \\ && 
- \half u (\e_3 \cdot \e_4)  (\e_1 \cdot k_4) ( \e_2 \cdot k_3)  
-  \half t  (\e_3 \cdot \e_4)  (\e_1 \cdot k_3) (\e_2 \cdot k_4) 
\eea
Factoring out the polarization vectors (taking into account that
they anti-commute with one another), one gets
\bea
\W_4 ^{(+)} 
\, \to \, - \e_1 ^\mu \e_2 ^\nu \e_3 ^\rho \e_4 ^\sigma
L_{\mu \nu \rho \sigma}
\eea
with 
\bea
L_{\mu \nu \rho \sigma}
 & = &
- {1 \over 4} t u \, \eta _{\mu \nu} \eta _{\rho \sigma} 
- {1 \over 4} st \, \eta _{\mu \rho}  \eta _{\nu \sigma}  
- {1 \over 4} su \, \eta_{\mu \sigma} \eta _{\nu \rho} 
\no \\ &&
+ \half s  \left ( 
    \eta _{\mu \rho} \, k_3^\nu k_1^\sigma 
    + \eta_{\mu \sigma}  \, k_4^\nu  k_1^\rho  
    + \eta _{\nu \rho}  \, k_3^\mu  k_2^\sigma 
    + \eta _{\nu \sigma} \,  k_4^\mu  k_2^\rho \right )
\no \\ && 
+  \half t  \left (
     \eta _{\mu \nu}  \, k_1^\rho k_2^\sigma  
    + \eta _{\mu \rho} \, k_1^\nu k_3^\sigma 
    + \eta _{\nu \sigma}   \, k_2^\mu  k_4^\rho 
    + \eta _{\rho \sigma}   \, k_3^\mu  k_4^\nu  \right )
\no \\ && 
+  \half u  \left ( \eta_{\mu \sigma}  \, k_1^\nu k_4^\rho  
    +  \eta _{\mu \nu}  \, k_2^\rho  k_1^\sigma  
    + \eta _{\nu \rho}   \,  k_2^\mu  k_3^\sigma  
    + \eta _{\rho \sigma}  \, k_4^\mu k_3^\nu   \right )
\eea
To identify this expression with the $K$-factor normalized earlier,
we identify the coefficients in $\e_1 \cdot \e_2$, and we find,
\bea 
K = (\e_1 \cdot \e_2) \left \{
2tu (\e_3 \cdot e_4) - 4 t (\e_3 \cdot k_1) (\e_4 \cdot k_2) \right \}
+ {\rm perm}
\eea
Hence, $ K = 8 \e_1 ^\mu \e_2 ^\nu \e_3 ^\rho \e_4 ^\sigma
L_{\mu \nu \rho \sigma}$ and 
\bea
\left \< \prod _{i=1} ^4 V(\e _i , \bar \e_i, k_i) \right \>
=
2^{-6} \left ( \int |\delta \rho _s|^2 \right ) K \tilde K 
 { 2 \pi \Gamma (-s/2) \Gamma (-t/2) \Gamma (-u/2)
\over 
\Gamma (1+s/2) \Gamma (1+t/2) \Gamma (1+u/2)}
\eea
Thus, the 4-point function for 4 gravitons is given by (\ref{tree4}) with 
$C_0 = 2^{-6}$.

\newpage

\section{One-loop superstring amplitudes}
\setcounter{equation}{0}

We first give a detailed derivation of the determinant formulas
for genus 1, paying special attention to their absolute normalization.

\subsection{Summary of determinant formulas}

Although the full amplitude is independent of the specific parametrizations
used, the intermediate formulas are dependent. We use the following
notations,
\bea
\Sigma & = &
\{ z \in {\bf C}, ~ z= \sigma _1 + \tau \sigma _2, \, 0\leq \sigma _{1,2} \leq 1 \}
\no \\
\M_1 & = &
\{ \tau \in {\bf C}, ~ 0 < \Im (\tau), ~  |\Re(\tau) | \leq \half , ~ 1 \leq |\tau|  \}
\eea
and the metric is given by $ds^2  = 2g_{z\bar z} dzd\bar z =2 dzd\bar z$.
With this metric, the area of the surface $\Sigma$ equals $2 \Im (\tau)$.
It is often convenient to set $z=x+iy$, with $x,y \in {\bf C}$.
The Cauchy-Riemann operators are then,
\bea
2\p = 2 {\p \over \p z} =  \p_x -i \p_y
\hskip 1in
2\bar \p = 2 {\p \over \p \bar z} = \p_x + i \p_y
\eea
With these conventions, we have the following values for the
functional determinants,
\bea
\det ' \Delta & = & 2 (\Im \, \tau ) ^2 |\eta (\tau) |^4
\no \\
(\det ' P_1 ^\dagger P_1)^\half & = & \det ' 2\Delta = (\Im \, \tau ) ^2 |\eta (\tau) |^4
\no \\
(\det 2 \bar \p )_\delta & = & (\det \bar \p) _\delta
= { \tet [\delta ](\tau) \over \eta (\tau)}
\no \\
( \det P_\half ^\dagger P_\half )^\half _\delta  & = & |( \det 2\p ) _\delta |^2
= \left | {\tet [\delta ](\tau) \over \eta (\tau)} \right |^2
\eea

\subsection{Determinants of Laplacians on the torus}

The differential operators are all proportional to the Cauchy-Riemann
operators or their associated Laplacian, but the boundary conditions
depend on whether we deal with a tensorial or a spinorial field
with spin structure $\delta$. The general boundary conditions are,
\bea
\f  (x+1 ,y) & = &  - \exp \{ 2 \pi i \delta ' \} \f  (x,y)
\no \\
\f  (x + \Re(\tau)  ,y + \Im (\tau) ) & = &  - \exp \{ 2 \pi i \delta '' \} \f  (x,y)
\eea
The Laplace operators for various tensor weights are all proportional
to the Laplace operator on scalars, $\Delta _{(0)} = -2 \p \bar \p$.
A basis of functions for these boundary conditions is given by
\bea
\f _{mn} [\delta ] (x,y) =
\exp \left \{ {2 \pi i  \over \Im ( \tau )}
\bigg ( (m+a) (\Im ( \tau) x - \Re( \tau) y) + (n+b) y  \bigg ) \right \}
\qquad m,n \in {\bf Z}
\eea
with $a =  \delta  ' + 1/2$ and $b = \delta  '' + 1/2$.
This basis diagonalizes the Laplace operators,
\bea
2 \Delta \f_{mn} [\delta ]
=
{ 4 \pi ^2 \over (\Im \, \tau)  ^2} \big | n+b - (m+a) \tau \big |^2 \f_{mn} [\delta ]
\eea
For generic $(a,b) \not= (0,0)$, the $\zeta$-function for the
Laplace operator is  defined  by
\bea
\zeta_\delta   (s) =
\sum _{m,n \in {\bf Z} }
{1 \over |n+b - (m+a) \tau | ^{2s}}
\eea
The determinant is obtained by
\bea
\ln ({\rm det}  2 \Delta)_\delta 
= - \zeta_\delta   '(0) - 2 \zeta _\delta   (0) \ln \left ( {\Im \, \tau \over 2 \pi} \right )
\eea
The values of the $\zeta$-function are as follows. As long as $(a,b) \not= (0,0)$,
we have  $\zeta _\delta (0) = 0$, and
\bea
- \zeta _\delta  '(0) = \ln \bigg | { \tet [\delta ](0,\tau) \over \eta (\tau)} \bigg |^2
\hskip 1in 
({\rm det}  2 \Delta )_\delta = \bigg | { \tet [\delta ](0,\tau) \over \eta (\tau)} \bigg |^2
\eea
and by holomorphic factorization, we have (up to an overall constant phase),
\bea
({\rm det}  2 \bar \p )_\delta = { \tet [\delta ](0,\tau) \over \eta (\tau)}
\eea
For periodic boundary conditions, $a=b=0$, we first
subtract the term $m=n=0$ in the case $(a,b) \not= (0,0)$ and
then take the limit $a,b \to 0$.  The relevant $\zeta$-function is
\bea
\zeta   (s) =
\sum _{(m,n) \not= (0,0) } {1 \over |n - m \tau | ^{2s}}
=
 \lim _{(a,b) \to (0,0)} \left ( \zeta _\delta (s) - {1 \over  |b-a\tau|^{2s}}  \right )
\eea
with 
\bea
\ln {\rm det} ' (2 \Delta )
=
- \zeta '(0) -2 \zeta (0) \ln \left ( {\Im \, \tau\over 2 \pi} \right )
\eea
Using $\tet [\delta ](0,\tau) \sim (b-a\tau) \tet _1 '(0,\tau) $, and 
$\tet _1 '(0,\tau) = -2 \pi \eta (\tau)^3$, we get,
\bea
{\rm det} ' (2 \Delta) = (\Im \, \tau ) ^2 |\eta (\tau)|^4
\eea
It is also useful to have the scaled up version of determinants,
\bea
{\rm det} ' (2\lambda \Delta) = {1 \over \lambda} (\Im \, \tau ) ^2 |\eta (\tau)|^4
\eea

\subsection{The 1-loop superstring measure}

Putting together the entire measure, we have
\bea
d\mu _1 [\delta]
=
\half \left ( { 4 \pi ^2 \alpha ' \det ' \Delta \over \int \sqrt{g} } \right )^{-5}
| ( \det \bar \p )_\delta |^{10}
(\det  P_\half ^\dagger P_\half )_\delta ^{ - \half }
\left ( { (\det ' P_1 ^\dagger P_1 )^\half \over 2 \Im \, \tau} \right )
\,  {|d\tau |^2 \over (\Im \, \tau )^2}
\eea
where the last factor $ |d\tau |^2 /\tau_2 ^2$ is the normalized Weil-Peterson
measure on $\M_1$. Substituting the values of the various determinants,
we have
\bea
d\mu _1 [\delta]
=
Q_1 { |\tet [\delta ](0, \tau)|^8
\over  (\Im \, \tau ) ^6 |\eta (\tau)|^{24}} |d\tau| ^2
\eea
where the coefficient $Q_1$ is given by
\bea
Q_1 = {1  \over 4 ( 4 \pi ^2\alpha ')^5}
\eea
A first factor of 1/2 arises  from dividing out by the volume factor of 
$SL(2,\bZ)$ instead of $PSL(2,\bZ)$, as has been argued for in
\cite{polchinski} and in \cite{dp86}.
The difference between these two groups consists of the 
(conformal) automorphism of the worldsheet $\Sigma$ which send $z\to -z$.
A second factor of 1/2 arises because of our conventions for 
$|d\tau|^2 = |d\tau \wedge d\bar \tau|$. Compared to \cite{dp88},
this has an extra factor of 2 by its very definition.

\medskip

As  a result of the  extra factor of 1/4 in $d\mu_1$ above, the coefficient 
$Q_1$ now also has an extra factor of 1/4. It is straightforward to see that, 
for the  superstring amplitude with even spin structure $\delta$, this 
final normalization  agrees with the corresponding normalization for the 
bosonic 1-loop amplitude, as computed in \cite{polchinski} and in \cite{dp86}, 
using the Polyakov integral, and in \cite{polchinskibook} using operator methods.

\subsection{The 1-loop superstring amplitudes}

To complete the calculation of the 4-point 1-loop amplitude,
we must carry out the GSO summation of the contractions
of all $\psi$-fields in the vertex operators. All other contractions
vanish by the Riemann identity on the torus. The expansion
of the vertex operator contraction, was given for any genus in
\cite{VI}, eq (7.2), with the normalization of the kinematical
factor $K$ given in \cite{VI}, eq (6.1). The two required
summation identities are as follows,
\bea
S_1 & = & \sum _\delta  \< \nu _0 |\delta \> \tet [\delta ](0)^4
S_\delta  (z_1,z_2)^2 S_\delta (z_3,z_4)^2
\no \\
S_2 & = & \sum _\delta  \< \nu _0 |\delta  \> \tet [\delta ](0)^4
S_\delta  (z_1,z_2) S_\delta  (z_2,z_3)  S_\delta  (z_3,z_4) S_\delta (z_4,z_1)
\eea
Since each quantity is holomorphic in each $z_i$, both are $z_i$-independent.
By setting $z_3 = z_1$ in the both, we obtain the same expressions;
hence we have $S_2=S_1$. To evaluate $S_1$, we use Fay's formula
for the torus,
\bea
S_\delta  (x,y)^2 =\p_x \p_y \ln E(x,y) + \tet [\delta ]''(0,\tau) /\tet [\delta ](0,\tau)
\eea
The first term does not contribute and the second may be recast
in the following form using the heat equation,
\bea
\tet [\delta  ]''(0,\tau) /\tet [\delta ](0,\tau) = 4\pi i \p_\tau \ln \tet [\delta  ](0,\tau)
\eea
Thus, we have
\bea
S_1 = (4 \pi i)^2 \sum _\delta  \< \nu _0 |\delta  \> \tet [\delta ](0)^4
\left ( \p_\tau \ln \tet [\delta  ](0,\tau) \right )^2
\eea
This sum may be worked out using the following identities,
based on \cite{IV}, eq (5.36-37),
\bea
\p _\tau \ln \left ( \tet [\mu_2] (0,\tau ) /\eta (\tau) \right )
& = &
{ i \pi \over 12} \left ( \tet [\mu _3](0,\tau)^4 + \tet [\mu _4](0,\tau)^4 \right )
\no \\
\p _\tau \ln \left ( \tet [\mu_3] (0,\tau ) /\eta (\tau) \right )
& = &
{ i \pi \over 12} \left ( \tet [\mu _2](0,\tau)^4 - \tet [\mu _4](0,\tau)^4 \right )
\no \\
\p _\tau \ln \left ( \tet [\mu_4] (0,\tau ) /\eta (\tau) \right )
& = &
{ i \pi \over 12} \left (- \tet [\mu _2](0,\tau)^4 - \tet [\mu _3](0,\tau)^4 \right )
\eea
where, as usual, we have 
\bea
\mu_2 = [\half ~ 0] \hskip .4in
\mu_3 = [0 ~ 0] \hskip .4in
\mu_4 = [0 ~ \half] 
\eea
Hence
\bea
S_1 & = &  {\pi ^4 \over 9} \bigg  [
- \tet [\mu_2]^4 (\tet [\mu_3]^4 + \tet [\mu_4]^4)^2
+ \tet [\mu_3]^4 (\tet [\mu_2]^4 - \tet [\mu_4]^4) ^2
\no \\ && \hskip .4in
- \tet [\mu_4]^4 (\tet [\mu_2]^4 + \tet [\mu_3]^4) ^2  \bigg ]
\eea
With the help of the Jacobi identity, this simplifies to
$S_1= - (2 \pi)^4 \eta (\tau)^{12}$.
Thus, using the notations of \cite{VI}, eq (7.2), we have
\bea
\sum _\mu \< \nu _0 |\mu \> \tet [\mu](0)^4 \W_0 [\mu]
=
- 4 \pi ^4 \eta (\tau)^{12} K
\eea
And the full 1-loop amplitude is given by
(\ref{one4}), with the value of $C_1$,
\bea
C_1 = (4 \pi ^4 )^2 Q_1 = {1 \over 2^8 \pi ^2 (\alpha ')^5}.
\eea

\newpage
\section{Regularization Dependence of Determinants}
\setcounter{equation}{0}
The Polyakov integral can only give superstring amplitudes up
to a proportionality factor of the form $e^{c (2h-2)}$, which 
should be absorbed in a shift of the dilaton expectation value. This is due to
the need for regularization and renormalization of the path integrals, and different
schemes differ by such a proportionality factor.

\subsection{Difference between regularization schemes}

Here we illustrate this phenomenon by comparing explicitly
the zeta function regularization with the short time cut-off regularization.

\medskip

In the zeta function regularization for the determinant of a Laplacian 
$\Delta$, the quantity $\ln \,{\rm det}'\,\Delta$ can be defined as 
$-\zeta'(0)$, where $\zeta(s)=\sum'\lambda^{-s}$, and the prime denotes summation over the non-zero eigenvalues $\lambda$ of $\Delta$.
For our purposes, it is convenient to write $\zeta(s)$ as
\be
\zeta(s)={1\over \Gamma(s)}\int_0^\infty dt \, t^{s-1} ({\rm Tr}\,e^{-t\Delta}-N)
\ee
where $N$ is the number of zero modes of $\Delta$.
Now the integral in $t$ between 1 and $\infty$ is convergent, and
produces an entire function of $s$. Thus, to evaluate $\zeta'(0)$, it suffices to determine explicitly the analytic continuation of the integral over $t\in (0,1)$. For this, write
\bea
\label{subtrac}
\int_0^1({\rm Tr}\,e^{-t\Delta}-N)t^{s-1}dt
=
\int_0^1({A_{-1}\over t}+A_0)t^{s-1}dt
+
\int_0^1({\rm Tr}\,e^{-t\Delta}-{A_{-1}\over t}-(A_0+N))t^{s-1}dt ~
\eea
where ${\rm Tr}\,e^{-t\Delta}-\sum_{i=-1}^NA_it^i={\cal O}(t^{N+1})$ is the asymptotic expansion of the trace of the heat kernel for small time. The second integral on the right is holomorphic at $s=0$. The first integral can be evaluated
explicitly to be $A_{-1}(s-1)^{-1}+A_0s^{-1}$ for ${\rm Re}\,s>1$, and hence it is given by this same formula by analytic
continuation. Altogether, we find
\bea
\ln\,{\rm det}'\Delta
&=&
A_{-1}-A_0({1\over s\Gamma(s)})'(0)
-
\int_0^1{dt\over t} ({\rm Tr}\,e^{-t\Delta}-{A_{-1}\over t}-(A_0+N))
\nonumber\\
&&
\qquad
-
\int_1^\infty{dt\over t}({\rm Tr}\,e^{-t\Delta}-N).
\eea 
We compare this value with the result of small time cut-off regularization and renormalization, ${\rm det}^*\Delta$, which is defined by
\be
\ln {\rm det}^*\Delta
=
-P.V.\int_{\epsilon}^\infty {dt\over t}({\rm Tr}\,e^{-t\Delta}-N),
\ee
where $P.V.$ indicates taking the finite part in the expansion in $\epsilon$ 
for small $\epsilon$. Once again, the integral over
$1\leq t<\infty$ converges, and we need only consider the integral over $\epsilon<t\leq 1$. Carrying out the rearrangement as in (\ref{subtrac}),
but with lower limit $\e$  and  $s=0$,
we find
\bea
\int_\epsilon^1{dt\over t}({\rm Tr}\,e^{-t\Delta}-N)
&=&-{A_{-1}\over\epsilon}+(A_0+N)\ln\,\epsilon
\nonumber\\
&&
\quad
+A_{-1}
-
\int_0^1{dt\over t}({\rm Tr}\,e^{-t\Delta}-{A_{-1}\over t}-(A_0+N)),
\eea 
and hence
\bea
\ln\,{\rm det}'\Delta = 
\ln\,{\rm det}^*\Delta -A_0 \left ({1\over s\Gamma(s)} \right )'(0)
\eea
Thus the two regularization schemes differ by an additive factor 
proportional to $A_0=\zeta(0)$. However, when $\Delta$ is the Laplacian 
on a field of U(1)  weight $n$, dimensional considerations show 
readily that the $t^0$ term of the heat kernel 
 must be proportional to the curvature 
of $\Sigma$. The coefficient $A_0$ is the integral
of this term, and is proportional to the Euler characteristic $\chi(\Sigma)=2-2h$.
This establishes the desired form for the difference between
regularization and renormalization schemes.

\subsection{Failure of ultra-locality}

The appearance of the factor $e^{c(2-2h)}$ is also related to
the failure of ultra-locality. Ultra-locality states that \cite{polchinski, weis},
\bea
\int \D x^\mu e^{- \lambda \pi  ||x^\mu ||_g ^2 } = e^{- \mu (\lambda) \int _\Sigma \sqrt{g}}
\eea
where $\lambda$ is a real positive constant, $\mu(\lambda)$ is a 
function of $\lambda $,  $||x^\mu ||_g$ is the $L^2$-norm 
for the worldsheet metric $g$ and $\D x^\mu$ the associated functional 
measure. Explicit calculation of the above functional integral, however,
shows that a term proportional to the Euler number also arises,
in full accord with standard renormalization theory, so that ultra-locality
does not hold when $h \not= 1$. This was already pointed out 
by Weisberger \cite{weis}.

\medskip

To see this explicitly, we calculate the integral in two different ways, 
first by bringing out the Laplacian $\Delta _g$, and then by bringing 
out the operator $\lambda \Delta _g$. Their equality gives
\bea
 \left ( { {\rm det} ' \Delta _g \over \int \sqrt{g} } \right )^{-d/2} 
\int dx^\mu _0 \int \D x^\mu e^{ - \pi \lambda ||x||_g^2}
 = \left ( { {\rm det} ' \lambda \Delta _g \over \int \sqrt{g} } \right )^{-d/2} 
\int dx^\mu _0 \int \D x^\mu e^{ - \pi ||x||_g^2}
\eea
Using now specifically zeta regularization for the determinants,
we have
\bea
({\rm det}' \lambda \Delta _g)^{-d/2} = \lambda ^{-\zeta _{\Delta _g} (0) d/2}
({\rm det}'   \Delta _g)^{-d/2},
\eea
and hence
\bea
\int \D x^\mu e^{- \lambda \pi  ||x^\mu ||_g ^2 } 
= \lambda ^{-\zeta _{\Delta _g} (0) d/2}
\int \D x^\mu e^{- \pi  ||x^\mu ||_g ^2 }. 
\eea
Since the expression $\zeta _{\Delta _g}(0)$ is proportional to the Euler number, our claim follows.

\end{appendix}

\end{document}